\documentclass[aps,pra,nofootinbib,twocolumn,floatfix,showpacs,superscriptaddress]{revtex4-1}
\usepackage[utf8]{inputenc}
\usepackage{amsmath,graphicx,epsfig,amsthm,color,bm}
\usepackage{pifont,bbding,amssymb,soul,mathrsfs}
\usepackage{indentfirst}
\usepackage{times,dsfont,units}
\usepackage{appendix, comment}
\usepackage{array}
\usepackage{makecell}
\usepackage{booktabs}
\usepackage{diagbox}
\usepackage{threeparttable}

\usepackage{tikz,pgfplots}

\newcommand{\tr}{{\rm Tr}}
\newcommand{\pr}{{\rm Pr}}

\newcommand{\liou}[1]{\ket{#1}\!\rangle}
\newcommand{\braliou}[1]{\langle\!\bra{#1}}
\newcommand{\braketliou}[2]{\langle\!\bra{#1}#2\rangle\!\rangle}

\newcommand{\bra}[1]{\langle #1|}
\newcommand{\ket}[1]{|#1\rangle}

\newcommand{\comments}[1]{}
\usepackage{lmodern}
\usepackage[T1]{fontenc}
\usepackage[hyperindex,breaklinks,colorlinks=true,citecolor=blue,urlcolor=blue]{hyperref}

\usepackage[capitalize]{cleveref}

\pgfplotsset{compat=1.17} 
\begin{document}
\title{Efficient Characterizations of Multiphoton States with an Ultra-thin Optical Device}
\author{Kui An}
\email{These authors contributed equally to this work.}
\affiliation{School of Physics, State Key Laboratory of Crystal Materials, Shandong University, Jinan 250100, China.}

\author{Zilei Liu}
\email{These authors contributed equally to this work.}
\affiliation{State Key Laboratory of Transient Optics and Photonics, Xi'an Institute of Optics and Precision Mechanics, Chinese Academy of Sciences, Xi'an 710119, China.}
\affiliation{University of Chinese Academy of Sciences, Beijing 100049,China.}

\author{Ting Zhang}
\affiliation{School of Physics, State Key Laboratory of Crystal Materials, Shandong University, Jinan 250100, China.}

\author{Siqi Li}
\affiliation{State Key Laboratory of Transient Optics and Photonics, Xi'an Institute of Optics and Precision Mechanics, Chinese Academy of Sciences, Xi'an 710119, China.}

\author{You Zhou}
\affiliation{Key Laboratory for Information Science of Electromagnetic Waves (Ministry of Education), Fudan University, Shanghai 200433, China}
\affiliation{Hefei National Laboratory, Hefei 230088, China}

\author{Xiao Yuan}
\affiliation{Center on Frontiers of Computing Studies, Peking University, Beijing 100871, China.}

\author{Leiran Wang}
\affiliation{State Key Laboratory of Transient Optics and Photonics, Xi'an Institute of Optics and Precision Mechanics, Chinese Academy of Sciences, Xi'an 710119, China.}
\affiliation{University of Chinese Academy of Sciences, Beijing 100049,China.}

\author{Wenfu Zhang}
\affiliation{State Key Laboratory of Transient Optics and Photonics, Xi'an Institute of Optics and Precision Mechanics, Chinese Academy of Sciences, Xi'an 710119, China.}
\affiliation{University of Chinese Academy of Sciences, Beijing 100049,China.}

\author{Guoxi Wang}
\email{wangguoxi@opt.ac.cn}
\affiliation{State Key Laboratory of Transient Optics and Photonics, Xi'an Institute of Optics and Precision Mechanics, Chinese Academy of Sciences, Xi'an 710119, China.}
\affiliation{University of Chinese Academy of Sciences, Beijing 100049,China.}

\author{He Lu}
\email{luhe@sdu.edu.cn}
\affiliation{School of Physics, State Key Laboratory of Crystal Materials, Shandong University, Jinan 250100, China.}
\affiliation{Shenzhen Research Institute of Shandong University, Shenzhen 518057, China.}

\begin{abstract}
Metasurface enables the generation and manipulation of multiphoton entanglement with flat optics, providing a more efficient platform for large-scale photonic quantum information processing. Here, we show that a single metasurface optical device would allow more efficient characterizations of multiphoton entangled states, such as shadow tomography, which generally requires fast and complicated control of optical setups to perform information-complete measurements, a demanding task using conventional optics. The compact and stable device here allows implementations of general positive observable value measures with a reduced sample complexity and significantly alleviates the experimental complexity to implement shadow tomography. Integrating self-learning and calibration algorithms, we observe notable advantages in the reconstruction of multiphoton entanglement, including using fewer measurements, having higher accuracy, and being robust against experimental imperfections. Our work unveils the feasibility of metasurface as a favorable integrated optical device for efficient characterization of multiphoton entanglement, and sheds light on scalable photonic quantum technologies with ultra-thin optical devices.
\end{abstract}

\maketitle
\section*{Introduction}
Metasurface, an ultra-thin and highly integrated optical device, is capable of full light control and thus provides novel applications in quantum photonics~\cite{Solntsev2021NPho}. In photonic quantum information processing, multiphoton entanglement is the building block for wide range of tasks, such as quantum computation~\cite{Walther2005Nature}, quantum error correction~\cite{Yao2012Nature}, quantum secret sharing~\cite{Bell2014NC,Lu2016PRL}, and quantum sensing~\cite{Liu2021NPhoton}. Recent investigations highlighted the feasibility of metasurface in generation~\cite{Li2020Science,Santiago2022Science},  manipulation~\cite{Georgi2019Light,Li2021NPho,Zhang2022Light}, and detection~\cite{Wang2018ScienceMeta,Wang2022APL} of multiphoton entanglement, indicating metasurface as a promising technology of ultra-thin optical device for large-scale quantum information processing.  

\begin{figure*}[ht!]
\centering
\includegraphics[width=\linewidth]{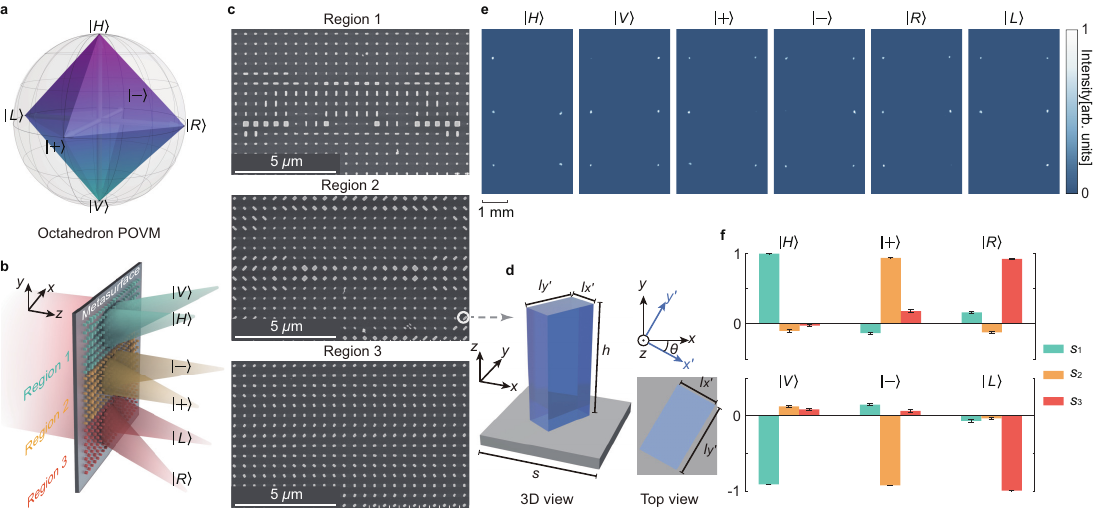}
\caption{\textbf{The metasurface-enabled octahedron positive operator valued measure~(POVM) $\mathbf E_\text{octa}$. a,} The elements in $\mathbf E_\text{octa}$ are projectors on states $\ket{H}$, $\ket{V}$, $\ket{+}$, $\ket{-}$, $\ket{R}$ and $\ket{L}$ respectively, which form a symmetric polytope of an octahedron on Bloch sphere. \textbf{b,} The metasurface to realize $\mathbf E_\text{octa}$, green, yellow and red blocks on the metasurface represent nanopillars with different arrangements. \textbf{c,} The scanning electron microscopy images of the fabricated nanopillars in three regions. \textbf{d,} Schematic drawing of single nanopillar that is fabricated with same height of 700nm but different $(\theta, l_{x^\prime}, l_{y\prime})$. \textbf{e,} The measured distribution of intensity on focal plane with input polarization of $\ket{H}$, $\ket{V}$, $\ket{+}$, $\ket{-}$, $\ket{R}$ and $\ket{L}$, respectively. \textbf{f}, The reconstructed Stokes parameters $(s_1, s_2, s_3)$ from data collected in \textbf{e}, and the error bars indicate standard deviations of reconstructed Stokes parameters.}
\label{Fig:expsetup}
\end{figure*} 

Characterization of multiphoton entanglement provides diagnostic information on experimental imperfections and benchmarks our technological progress towards the reliable control of large-scale photons. The standard quantum tomography~(SQT)~\cite{James2001PRA} requires an exponential overhead with respect to the system size. Recently, more efficient protocols have been proposed and demonstrated with fewer measurements, such as compressed sensing~\cite{Gross2010PRL, Liu2018PRL}, adaptive tomography~\cite{Mahler2013PRL,Hou2016npjqi,Granade2017NJP} and self-guided quantum tomography~(SGQT)~\cite{Ferrie2014PRL,Chapman2016PRL,Rambach2021PRL}. Shadow tomography, which was first proposed by Aaronson~\emph{et al.}~\cite{Aaronson2018shadow} and then concreted by Huang~\emph{et al.}~\cite{Huang2020NPhysics}, efficiently predicts functions of the quantum states instead of state reconstruction. Huang's protocol~\cite{Huang2020NPhysics} is hereafter referred as shadow tomography. Shadow tomography is efficient in estimation of quantities in terms of observable~(polynomial), including nonlinear observables such as purity and R$\acute{\text{e}}$nyi entropy~\cite{Brydges2019Science,elben2020cross,garcia2021quantum,Zhenhuan2022correlation}, which is of particular interest in detecting multipartite entanglement ~\cite{Elben2020PRL,singlezhou,Neven2021npjqi,Zhang2021PRL} and thus is helpful in benchmarking the technologies towards generation of genuine multipartite  entanglement~\cite{Lu2018PRX,zhou2019detecting,zhou2022scheme}. Nevertheless, shadow tomography generally requires the experimental capability of performing information-complete measurements, leading to the consequence that the time of switching experimental setting is much longer than that of data acquisition. A potential solution is to replace the unitary operations and projective measurements with positive operator valued measures (POVMs), which is capable to extract complete information in a single experimental setting~\cite{Acharya2021PRA,Stricker2022experimental,Nguyen2022PRL}. The POVM significantly alleviates the experimental complexity to perform shadow tomography, and thus enables the real-time shadow tomography, i.e., an experimentalist is free to stop shadow tomography at any time. However, a compact and scalable implementation of POVM in optical system is still technically challenging. On the other hand, shadow tomography is not able to easily predict the properties that cannot be directly expressed in terms of observables~(polynomial) such as von Neumann entropy $S(\rho)=-\tr(\rho \log \rho)$, which is key ingredient in topological entanglement entropy~\cite{Kitaev2006PRL,Satzinger2021science}. 

In this work, we report an implementation of POVM enabled by a metasurface, which is based on planar arrays of nanopillars and able to provide complete control of polarization. The POVM we achieved allows to implement real-time shadow tomography, and observe the shadow norm that determines sample complexity. Moreover, we show that the metasurface-enabled shadow tomography can be readily equipped with other algorithms, enabling the unexplored advantages of shadow tomography . In particular, we propose and implement shadow tomography optimized by simultaneous perturbation stochastic approximation~(SPSA)~\cite{Spall1992}, the so-called self-learning shadow tomography~(SLST). SLST efficiently returns a physical state with high accuracy against the metasurface-induced imperfections, which can be further used to calculate the quantities that cannot be expressed in terms of directly observable. We also implement robust shadow tomography~\cite{Chen2021PRXQuantum} to show the robustness of reconstruction against the engineered optical loss.

\begin{figure*}[t]
\centering
\includegraphics[width=\linewidth]{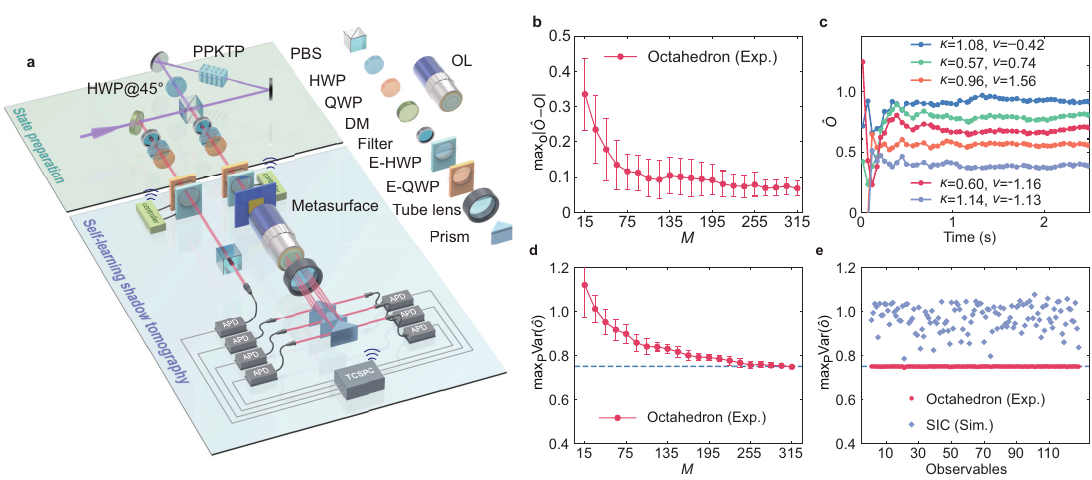}
\caption{\textbf{The experimental setup and results of shadow tomography with metasurface-enabled positive operator valued measure~(POVM). a,} Setup to generate entangled photons and demonstrate shadow tomography with metasurface. PBS: polarizing beam splitter. DM: dichroic mirror. HWP: half-wave plate. QWP: quarter-wave plate. E-HWP: electrically-rotated HWP. E-QWP: electrically-rotated QWP. OL: objective lens. \textbf{b,} The maximal error in estimation of expectation of $O\in\mathbf{O}$. \textbf{c,} The real-time estimation of expectation of five randomly selected $O\in\mathbf{O}$. \textbf{d,} The results of shadow norm $\max_\mathbf{P}\text{Var}(\hat{o})$ for $O=\ket{+}\bra{+}$ with different experimental runs. \textbf{e,} The results of shadow norm for 128 $O\in\mathbf{O}$~(red dots), and the simulated results of shadow norm with symmetric informationally complete~(SIC) POVM~(blue diamonds). The 128 observables $O\in\mathbf O$ are selected according to Haar random. The dots and bars in~\textbf{b} and~\textbf{d} are the mean value and corresponding standard deviations obtained by repeating the experiment 5 times. The abbreviations of Exp. and Sim. indicate experimental results and simulation results respectively.}
\label{Fig:datavariance}
\end{figure*}

\section*{Results}
\textbf{Shadow tomography with POVM. }We start by briefly reviewing the shadow tomography with POVM. Considering a 2-level (qubit) quantum system, a set of $L$ rank-one projectors $\{\ket{\psi_l}\bra{\psi_l}\in\mathbb{H}_2\}_{l=1}^{L}$ is called a quantum 2-design if the average value of the second-moment operator $(\ket{\psi_l}\bra{\psi_l})^{\otimes 2}$ over the set is proportional to the projector onto the totally symmetric subspace of two copies~\cite{Guta2020}. Each quantum 2-design is proportional to a POVM $\mathbf E=\{\frac{2}{L}\ket{\psi_l}\bra{\psi_l}\}_{l=1}^{L}$
with the element $E_l=\frac{2}{L}\ket{\psi_l}\bra{\psi_l}$ being positive semidefinite and satisfying $\sum_{l=1}^{L}E_l=\mathds 1_2$. Note that quantum 1-design is sufficient to form a POVM but is not always information-complete for tomography, such as the measurement on computational basis $\{\ket{0},\ket{1}\}$. Measuring a quantum state $\rho$ using POVM $\mathbf E$ results one $l\in[L]$ outcome with probability $\pr(l|\rho)=\tr(E_l\rho)$ according to Born's rule. The POVM~$\mathbf E$ together with the preparation of the corresponding state $\ket{\psi_l}$ can be viewed as a linear map $\mathcal M: \mathbb{H}_2\to\mathbb{H}_2$, and the `classical shadow' is the solution of least-square estimator with single experimental run,
\begin{equation}
\hat{\rho}^{(m)}_l=\mathcal M^{-1}(\ket{\psi_l}\bra{\psi_l})=3\ket{\psi_l}\bra{\psi_l}-\mathds1_2.
\end{equation}
For an $N$-qubit state, the classical shadow is the tensor product of simultaneous single-qubit estimations $\hat{\rho}^{(m)}=\bigotimes_{n=1}^N\hat{\rho}^{(m)}_{l_n}$ with $l_n$ being the outcome of $n$-th qubit, and one has $\mathbb{E}[\hat{\rho}^{(m)}]=\rho$. Repeating the POVM $M$ times~(experimental runs), one has a collection of classical shadows $\{\hat{\rho}^{(m)}\}_{m=1}^M$, which is further inquired for estimation of various properties of the underlying state. See Supplementary Note 1A for more details.

\begin{figure*}[t]
\centering
\includegraphics[width=\linewidth]{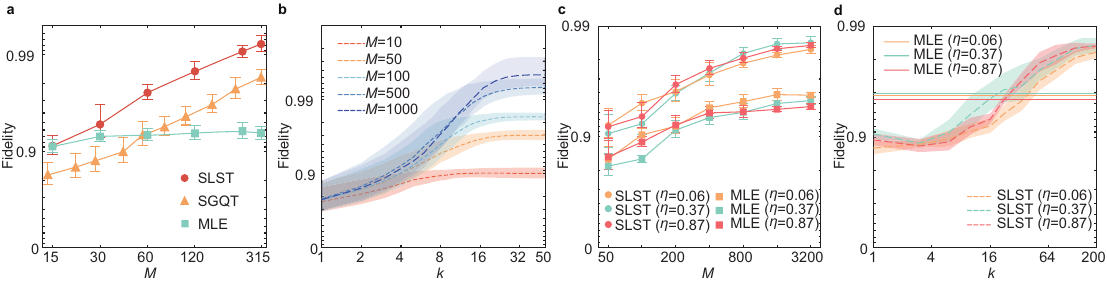}
\caption{\textbf{Experimental results of self-learning shadow tomography~(SLST) on one-photon and two-photon states. a,} The average fidelity between reconstructed single-photon states $\tau$ and target state $\ket{\psi}_{\gamma,\phi}$ using SLST, self-guided quantum tomography~(SGQT), maximum likelihood estimation~(MLE) reconstruction. \textbf{b,} Average fidelity of SLST by increasing experimental runs $M$ from 10 to 1000. \textbf{c,} Fidelity between reconstructed two-photon states $\tau$ and target state $\rho_{\eta}$ using SLST and MLE. 
\textbf{d,} The fidelities of two-photon states reconstruction from SLST (dash lines) with $M=2000$ measurements. The solid lines represent the fidelity from MLE tomography with $M=2000$ measurements.  The dots and bars in~\textbf{a} and~\textbf{c} are the mean value and the corresponding standard deviations obtained by repeating the experiment 5 times. The dashed lines and shadings in~\textbf{d} and~\textbf{e} are the mean value and standard deviation obtained by repeating the iteration 5 times. }
\label{Fig:data}
\end{figure*}

\noindent\textbf{Implementation of POVM with metasurface. }In our experiment, we focus on the POVM on polarization-encoded qubit, i.e., $\ket{0(1)}=\ket{H(V)}$ with $\ket{H(V)}$ being the horizontal (vertical) polarization, and consider POVM of $L=6$ and $\ket{\psi_l}\in\{\ket{H},\ket{V}, \ket{+}, \ket{-}, \ket{R}, \ket{L}\}$ with $\ket{\pm}=(\ket{H}\pm\ket{V})/\sqrt{2}$ and $\ket{R(L)}=(\ket{H}\pm i\ket{V})/\sqrt{2}$. The corresponding POVM $\mathbf E_\text{octa}$ is described by a symmetric polytope of an octahedron on Bloch sphere as shown in Figure~\ref{Fig:expsetup}a. To realize $\mathbf E_\text{octa}$, we design and fabricate a 210$\mu$m$\times$210$\mu$m polarization-dependent metasurface that splits incident light into six directions corresponding to projection on $\ket{H}$, $\ket{V}$, $\ket{+}$, $\ket{-}$, $\ket{R}$ and $\ket{L}$ with equal probability (shown in Figure~\ref{Fig:expsetup}b). Note that projection on $\ket{\psi_l}$ with equal probability is guaranteed with post-selection to eliminate the mode mismatch between incident light~(Gaussian beam) and metasurface~(square)~(see Supplementary Note 5 for details). The metasurface is an array (with square pixel of $s=500$nm) of single-layer amorphous silicon nanopillars on quartz substrate as shown in Figure~\ref{Fig:expsetup}c and Figure~\ref{Fig:expsetup}d. The nanopillars are with the same height of $700$nm but different $l_{x^\prime}$, $l_{y^\prime}$ and orientation $\theta$ relative to the reference coordinate system. In this sense, a single nanopillar can be regarded as a waveguide with different rectangular cross profile that exhibits corresponding effective birefringence, leading to spatial separation between orthogonal polarizations~\cite{Arbabi2015NN}. The metasurface is divided into three regions with same size of $210\mu$m$\times$70$\mu$m but different arrangement of nanopillars, i.e., $(\theta, l_{x^\prime}, l_{y^\prime})$. By carefully designing the arrangement of nanopillars, we can realize spatial separation of $\ket{H}/\ket{V}$, $\ket{+}/\ket{-}$ and $\ket{R}/\ket{L}$, respectively. To validate the capability of fabricated metasurface to perform information-complete measurement, we test metasurface with input states of $\ket{\psi_l}$ and measure the distribution of output intensity on focal plane. The results are shown in Figure~\ref{Fig:expsetup}e, according to which we reconstruct the Stokes parameters $(s_1, s_2, s_3)$ shown in Figure~\ref{Fig:expsetup}f. Compared to the ideal values, the average errors of reconstructed Stokes parameters are $0.101\pm0.005$, $0.086\pm0.005$ and $0.073\pm0.005$, respectively. These errors are mainly caused by the discretization of phase front in design, which inevitably introduces higher-order deflections~\cite{Sell2017NL}~(see Supplementary Note 4 for details of metasurface).

\noindent\textbf{Estimation of observables. }We first perform shadow tomography with the fabricated metasurface on single-photon pure state $\ket{\psi_{\gamma,\phi}}=\cos\gamma\ket{H}+\sin\gamma\text{e}^{i\phi}\ket{V}$ with $\gamma=0.91$ and $\phi=0.12$. As shown in Figure~\ref{Fig:datavariance}a, the polarization-entangled photons (central wavelength of 810~nm) are generated from a periodically poled potassium titanyl phosphate (PPKTP) crystal placed in a Sagnac interferometer via spontaneous parametric down conversion~(SPDC), which is pumped by a laser diode (central wavelength of 405~nm). The generated entangled photons are with ideal form of $\ket{\psi}_\eta=\sqrt{\eta}\ket{HV}+\sqrt{1-\eta}\ket{VH}$, where $\eta$ is determined by polarization of pump light. Projecting one photon of $\ket{\psi}_\eta$ on $\ket{H}$ heralds the other photon on state $\ket{V}$, which can further be transformed to arbitrary $\ket{\psi_{\gamma,\phi}}=\cos\gamma\ket{H}+\sin\gamma\text{e}^{i\phi}\ket{V}$ by a combination of electrically-rotated half-wave plate~(E-HWP) and quarter-wave plate~(E-QWP). Then, the heralded photon passes through the metasurface, and is coupled to six multimode fibers at outputs using an objective lens~(OL), a tube lens, and three prisms, respectively. With the collection of classical shadows $\{\hat{\rho}^{(m)}\}_{m=1}^{M}$, we focus on the estimation of observables in set of 128 single-qubit projections, i.e., $O=\ket{\psi_{\kappa,\nu}}\bra{\psi_{\kappa,\nu}}\in\mathbf O$ with $\ket{\psi_{\kappa,\nu}}=\cos\kappa\ket{H}+\sin\kappa\text{e}^{i\nu}\ket{V}$ being uniformly distributed on Bloch sphere. The estimation of expected value of observable is $\hat{O}=1/M\sum_{m=1}^{M}\hat{o}^{(m)}$, where $\hat{o}^{(m)}=\tr(O\hat{\rho}^{(m)})$ is the i.i.d single-shot estimator. Note that $\hat{O}$ converges to the exact expectation value $\tr(\rho O)$ as $M\to\infty$. The error of estimation with metasurface-enabled POVM is indicated by the distance between $\hat{O}$ and ideal expectation $\langle O\rangle=\bra{\psi_{\gamma,\phi}}O\ket{\psi_{\gamma,\phi}}$. As shown in Figure~\ref{Fig:datavariance}b, the maximal distance $\max_\mathbf{O}\|\hat{O}-\langle O\rangle\|$ converges to 0.07 with the increase of $M$, which is consistent with the error we obtained in reconstruction of Stokes parameters. In Figure~\ref{Fig:datavariance}c, we show the real-time estimation of $\hat{O}$ by randomly selecting five $O\in\mathbf{O}$, in which we observe the convergence of $\hat{O}$ after a few hundreds of milliseconds.

The sample complexity of estimation is further characterized by the variance 
$\text{Var}(\hat{O}) = \text{Var}(\hat{o})\leq ||O||^2_\text{shd}$.
Here the shadow norm $||O||^2_\text{shd}$
~\cite{Huang2020NPhysics} is the maximization of $\text{Var}(\hat{o})$ over all possible states $\rho$ to remove the state-dependence. For ideal $\mathbf E_\text{octa}$, the shadow norm $||O||^2_\text{shd}=0.75$ regardless of the explicit form of $O\in\mathbf O$~(see Supplementary Note 1D for deviation of $||O||^2_\text{shd}=0.75$). Experimentally, the variance of a single-shot estimation is
\begin{equation}\label{Eq:varexp}
\text{Var}(\hat{o})=\frac{1}{M}\sum_{m=1}^{M} \left(\hat{o}^{(m)} - \hat{O} \right)^2.
\end{equation}
It is impossible to maximize $\text{Var}(\hat{o})$ over all possible $\ket{\psi_{\gamma,\phi}}$ in experiment, so that we prepare totally 20 $\ket{\psi_{\gamma,\phi}}$ that are uniformly distributed on Bloch sphere, forming a state set of $\mathbf{P}$ . For each prepared $\ket{\psi_{\gamma,\phi}}$, we perform shadow tomography and estimate the expectation of $O=\ket{+}\bra{+}$. The results of $\max_\mathbf{P}\text{Var}(\hat{o})$ are shown in Figure~\ref{Fig:datavariance}d, in which we observe that $\max_\mathbf{P}\text{Var}(\hat{o}^{(m)})$ converges to 0.75 when $M>255$. In Figure~\ref{Fig:datavariance}e, we show $\max_\mathbf{P}\text{Var}(\hat{o})$ of 128 observables $O\in \mathbf O$ with $M=315$ measurements, which agrees well with the theoretical prediction that the shadow norm is a constant regardless of the explicit form of $O\in\mathbf O$~\cite{Nguyen2022PRL}. To give a comparison, we simulate $\max_\mathbf{P}\text{Var}(\hat{o})$ with symmetric informationally complete~(SIC) POVM $\mathbf E_\text{SIC}$~\cite{Renes2004JMP}, which is constructed with the minimal number of 4 measurements for qubit system and has been widely adopted in investigations of advanced tomography~\cite{Torlai2018NP, Carrasquilla2019NMI, Garc2021PRXQ}. As shown in Figure~\ref{Fig:datavariance}e, the shadow norm with $\mathbf E_\text{SIC}$ depends on observable $O$ and generally larger than that with $\mathbf E_\text{octa}$, which indicates $\mathbf E_\text{octa}$ requires less shots $M$ than $\mathbf E_\text{SIC}$ to achieve the same accuracy of estimation $\hat{O}$.

\noindent\textbf{State reconstruction. } The direct estimation from classical shadows $\hat{\rho}^{(m)}$, i.e., $\hat{\rho}=1/M\sum_{m=1}^{M}\hat{\rho}^{(m)}$, is generally not a physical state with finite $M$ measurements, which limits the application of shadow tomography in estimation of nonlinear functions~\cite{Stricker2022experimental,zhou2022hybrid}. Physical constraints need to be introduced to enforce the positivity of the reconstructed state $\tau$, which can be addressed by solving the optimization problem 
\begin{widetext}  
\begin{equation}
\label{Eq:opt}
    \begin{split}
        \text{minimize}\hspace{1.2cm}&\hat{N}_F(\tau)=\frac{2}{M(M-1)}\sum_{m<n}\tr\left[\hat{\rho}^{(m)}\hat{\rho}^{(n)}\right]+ \tr(\tau^2)-2 \sum_{m}\tr(\hat{\rho}^{(m)}\tau)\\
        \text{subject to}\hspace{1.2cm}&\tau\geq0,\hspace{2pt}\tr(\tau)=1,
    \end{split}
\end{equation}
\end{widetext}
where $\tau$ is the proposed state that is positive semidefinite ($\tau\geq0$) with unit trace ($\tr(\tau)=1$), and the cost function $\hat{N}_F(\tau)$ is the unbiased estimator of squared Frobenius norm with $\{\hat{\rho}^{(m)}\}$~(see Supplementary Note 2A for more details). Note that the squared state fidelity adopted in SGQT~\cite{Ferrie2014PRL,Chapman2016PRL,Rambach2021PRL} is not an unbiased estimator with $\{\hat{\rho}^{(m)}\}$ for mixed state.
We employ an iterative self-learning algorithm, i.e., SPSA algorithm, to solve the optimization problem in Eq.~(\ref{Eq:opt}). SPSA is especially efficient in multi-parameter optimization problems in terms of providing a good solution for a relatively small number of measurements of the objective function~\cite{spall2005introduction}, which holds the similar spirit as shadow tomography. In traditional maximum likelihood estimation~(MLE) reconstruction~\cite{james2001measurement}, the computational expense required to estimate gradient direction is directly proportional to the number of unknown parameters~($4^N-1$ for an $N-$qubit state) as it approximates the gradient by varying one parameter at a time, which becomes an issue when the number of qubit is large. In SPSA, the minimization of cost function $\hat{N}_F(\tau)$ is achieved by perturbing  all parameters simultaneously, and one gradient evaluation requires only two evaluations of the cost function. While SPSA costs more iterations to converge, it returns state with higher fidelity in limited number of iterations compared to MLE~\cite{Chapman2016PRL}. More importantly, SPSA formally accommodates noisy measurements of the objective function, which is an important practical concern in experiment. 

Generally, an $N$-qubit state $\tau$ can be modeled with $d^2$ parameters with $d=2^N$ being the dimension of $\tau$. Thus, the proposed state $\tau$ is determined by a $d^2$-dimensional vector $\bm{r}=[r_1, r_2,\cdots, r_{d^2}]$. SPSA optimization estimates the gradient by simultaneously perturbing all parameters $r_i$ in a random direction, instead of individually addressing each $r_i$. In $k$-th iteration, the simultaneous perturbation approximation has all elements of $\bm{r}_k$ perturbed together by a random perturbation vector $\boldsymbol{\Delta}_k =[\Delta_{k1}, \Delta_{k2}, \cdots, \Delta_{kd^2}]$ with $\Delta_{ki}$ being generated from Bernoulli $\pm1$ distribution with equal probability. Then the gradient is calculated by
\begin{equation}\label{Eq:gradient}
\bm{g}_k = \frac{\hat{N}_F(\bm{r}_k+B_k\boldsymbol{\Delta}_k) - \hat{N}_F(\bm{r}_k-B_k\boldsymbol{\Delta}_k)}{2 B_k}\boldsymbol{\Delta}_k,
\end{equation}
and $\bm{r}_k$ is updated to $\bm{r}_{k+1}$ by $\bm{r}_{k+1} = \bm{r}_{k}+A_{k}\bm{g}_k$. $A_k$ and $B_k$ are functions in forms of $A_k = {a_1}/{(k+a_2)^{a_3}}$ and $B_k = {b_1}/{k^{b_2}}$ with $a_1$, $a_2$, $a_3$, $b_1$ and $b_2$ being hyperparameters that determine the convergence speed of algorithm, which can be generally obtained from numerical simulations~(see Supplementary Note 2B for hyperparameter settings). SLST is terminated when there is little change of $\hat{N}_F(\bm{r}_k)$ in several successive iterations, and corresponding $\tau_k$ is the reconstructed state. We emphasize that SPSA inevitably introduces systematic errors of the reconstructed state, as well as other optimization algorithms such as MLE and least squares~\cite{Schwemmer2015PRL}. In fact, it is a tradeoff that the reconstruction of a physical state suffers from a bias. 

\begin{figure*}[t]
\centering
\includegraphics[width=\linewidth]{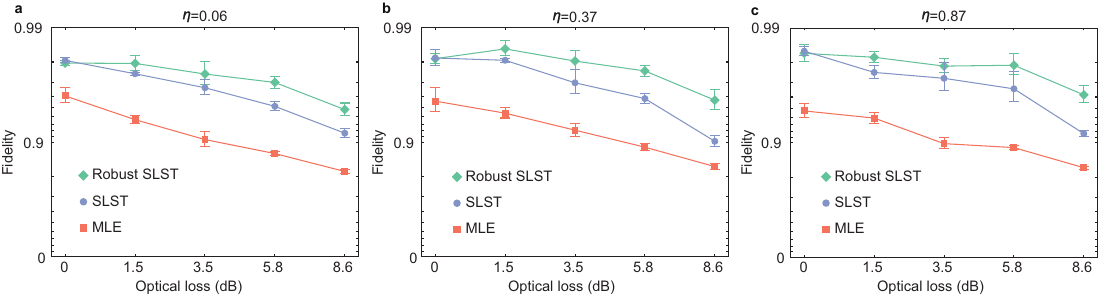}
\caption{\textbf{Results of fidelities from robust self-learning shadow tomography~(SLST), SLST and maximum likelihood estimation~(MLE) reconstruction on two-photon states a, }$\rho_{\eta=0.06}$, \textbf{b, }$\rho_{\eta=0.37}$ and \textbf{c, }$\rho_{\eta=0.87}$. In each reconstructions, the experiment is carried out with $M=1000$ runs. In robust SLST, additional $M^\prime=2000$ experimental runs are used for calibration. We set $k=200$ in robust SLST and SLST. The error bars are the standard deviations in SLST~(robust SLST), obtained from Monte Carlo simulation with assumption that the collected photons in $M$~($M^\prime$ and $M$) experimental runs have Poisson distribution.}
\label{Fig:datawithnoise}
\end{figure*}

As the prepared single-photon state is extremely closed to the ideal state $\ket{\psi_{\gamma,\phi}}$, the accuracy of reconstruction is characterized by the state fidelity between returned state $\tau_k$ and ideal state $\ket{\psi_{\gamma,\phi}}$, i.e., $F=\sqrt{\tr(\tau_k\ket{\psi_{\gamma,\phi}}\bra{\psi_{\gamma,\phi}})}$. The results of average fidelity of SLST over 20 prepared $\ket{\psi_{\gamma,\phi}}\in\mathbf{P}$ after $k=30$ iterations are shown with red dots in Figure~\ref{Fig:data}a, where the average fidelity increases as $M$ increases and achieves 0.992$\pm$0.001 with $M=315$ measurements. The fabricated metasurface is also capable to collect data required for state reconstruction with other technologies, i.e., SGQT~\cite{Ferrie2014PRL,Chapman2016PRL,Rambach2021PRL} and MLE reconstruction~(see Supplementary Note 3 for demonstration of SGQT). In SGQT, two projective measurements are performed with 7 experimental runs in each iteration, and SPSA is used to update the proposed state $\tau^{\text{SGQT}}$. The results of $F=\sqrt{\tr(\tau^{\text{SGQT}}\ket{\psi_{\gamma,\phi}}\bra{\psi_{\gamma,\phi}})}$ are shown with yellow triangles in Figure~\ref{Fig:data}a, in which we observe an average fidelity of 0.983$\pm$0.003 after 45 iterations~(total experimental runs of 315 as the same as that in SLST). The results of MLE reconstruction $F=\sqrt{\tr(\tau^{\text{MLE}}\ket{\psi_{\gamma,\phi}}\bra{\psi_{\gamma,\phi}})}$ are shown with cyan squares in Figure~\ref{Fig:data}a. When $M$ is small ($M<60$), MLE reconstruction is more accurate than SGQT. However, SLST always exhibits higher accuracy compared to other techniques with the same number of experimental runs. It is worth noting that the average fidelity with MLE reconstruction converges to $0.93\pm0.01$ and the error of reconstruction is about 0.07, which is consistent with errors in estimation of observables in Figure~\ref{Fig:datavariance}b. Although the error of metasurface reduces the accuracy of shadow tomography and MLE reconstruction, SLST and SGQT with SPSA optimization can dramatically suppress metasurface-induced error as SPSA can accommodate noisy measurements of the cost function. The accuracy of SLST does not keep increasing with the number of iterations as reflected in Figure~\ref{Fig:data}b, where the converged fidelity depends on the number of experimental runs $M$ in classical shadow collection. 

We also demonstrate SLST on two-photon entangled states $\ket{\psi}_\eta=\sqrt{\eta}\ket{HV}+\sqrt{1-\eta}\ket{VH}$ with $\eta=0.06, 0.37$ and $0.87$. In two-photon SLST, one photon is detected by metasurface-enabled $\mathbf E_\text{octa}$, and the other photon is detected by randomly choosing $\sigma_x$, $\sigma_y$ and $\sigma_z$ measurements realized by an E-HWP and an E-QWP. In contrast to the single-photon state, the generated two-photon state $\rho_\eta$ is far from pure state as it is affected by more noises that are mainly attributed to high-order emission in SPDC and mode mismatch when overlapping two photons in Sagnac interferometer. Thus, the proposed state $\tau_k$ should be a mixed state in general form of $\tau_k=T^\dagger T$ with $T$ being a complex lower triangular matrix~(see Supplementary Note 2C for details). Accordingly, the accuracy of reconstruction is characterized by the fidelity between returned state $\tau_k$ and $\rho_\eta$, where $\rho_\eta$ is MLE reconstruction with large amount of data~($M\approx 8\times10^5$) collected from bulk optical setting~(waveplates and PBS). The results of $F=\tr \left( \sqrt{ \sqrt{\tau_{200}} \rho_\eta \sqrt{\tau_{200}}} \right)$ are shown with dots in Figure~\ref{Fig:data}c, the fidelities of three states reach 0.986$\pm$0.002, 0.990$\pm$0.001 and 0.981$\pm $0.002 with $M=2000$ experimental runs and $k=200$ iterations. We also perform MLE reconstruction $\tau^\text{MLE}$ of two-photon states, where one photon is detected by metasurface and the other is detected by bulk optical setting. The results of $F=\tr \left( \sqrt{ \sqrt{\tau^\text{MLE}} \rho_\eta \sqrt{\tau^\text{MLE}}} \right)$ with $M$ experimental runs are shown with squares in Figure~\ref{Fig:data}c. The error in two-photon MLE reconstruction is about $0.047\pm0.005$, which is smaller than that in single-photon MLE reconstruction as only one photon is detected by noisy device~(metasurface). In Figure~\ref{Fig:data}d, we show that the fidelity of SLST with $M=2000$ is converging after $k=200$ iterations.

\noindent\textbf{Robust shadow tomography. }Finally, we demonstrate robustness of SLST can be further improved by robust shadow tomography~\cite{Chen2021PRXQuantum,Koh2022Quantum}. Considering that the measurement apparatus are noisy, the measurement apparatus can be calibrated prior to performing SLST. To this end, shadow tomography is firstly performed on high-fidelity state $\ket{HH}$ with $M^\prime$ experimental runs to calculate the noisy quantum channel $\widetilde{\mathcal M}$. Consequently, the classical shadow is constructed by the noisy channel, i.e., $\hat{\rho}^{(m)}=\widetilde{\mathcal M}^{-1}(\ket{\psi_{l_1}}\bra{\psi_{l_1}}\otimes\ket{\psi_{l_2}}\bra{\psi_{l_2}})$~(See Supplementary Note 1B for details of robust shadow tomography). The framework of robust shadow tomography is valid in our experimental setting. Firstly, although two photons are detected with different measurement devices, i.e., one is the metasurface-enabled POVM while the other is randomly detected on three Pauli bases, the mathematical models of these two measurement devices are identical. Secondly, although the metasurface-induced measurement errors are different between six projections, it has been shown that gate-dependent noise can be suppressed by robust shadow tomography~\cite{Chen2021PRXQuantum}. Finally, the experimental device is able to generate $\ket{HH}$ with sufficiently high fidelity. Otherwise, the noise in state preparation might be added in $\widetilde{\mathcal M}$, which introduces biased estimation of returned state. In our experiment, the fidelity of prepared $\ket{HH}$ is 0.9956$\pm$0.0005 with respect to the ideal form. To demonstrate robust SLST, we insert a tunable attenuator before metasurface to introduce optical loss from 1.5dB to 8.6dB, which accordingly reduces the fidelity of prepared state as reflected by the MLE reconstruction shown in Figure~\ref{Fig:datawithnoise}. Compared to SLST, robust SLST is able to enhance the accuracy of reconstruction in the presence of optical loss, especially at the high-level optical loss. It is worth mentioning that SLST itself can accommodate metasurface-induced measurement errors so that the enhancement of robust SLST is not significant when optical loss is zero. Increasing the optical loss is equivalent to stronger measurement noise. We observe the significant enhancement of robust SLST at high-level optical loss, which indicates robust SLST can further improve the robustness of SLST against noise~(See Supplementary Note 1C for numerical simulations of robust SLST).

\section*{Discussion}
We propose and demonstrate POVM with a single metasurface that enables implementation of real-time shadow tomography and observation of sample complexity. Together with the developed SLST, the underlying quantum states can be reconstructed efficiently, accurately and robustly. The advantages are evident even in single- and two-photon polarization-encoded states. The concept of octahedron POVM can be readily realized with integrated optics, where the directional couplers and phase shifters are able to construct octahedron POVM encoded in path degree of freedom. Metasurface-enabled POVM is particularly promising for efficient detection of scalable polarization-encoded multiphoton entanglement, in which two measurement devices are sufficient for full characterization~\cite{Evan2022PRL}. Our investigation is compatible with metasurface-enabled generation~\cite{Li2020Science,Santiago2022Science} and  manipulation~\cite{Georgi2019Light,Li2021NPho,Zhang2022Light} of photonic states, thereby opening the door to quantum information processing with a single ultra-thin optical device.

\section*{Methods}
\noindent\textbf{Fabrication of metasurface.} A 700~nm-thick layer of a-Si is deposited on top of 750~$\mu$m-thick fused quartz wafers using the low-pressure chemical vapor deposition (LPCVD) technique. Then a layer of AR-P6200.09 resists~(Allresist GmbH) with a thickness of 200 nm is spun and coated on the substrate. The metasurface pattern is generated with electron-beam lithography (EBL) process which is set with 120~kV, 1~nA current and 300~$ \mu$c $\text{cm}^{-2}$ dose. Subsequently, the resist is developed with AR300-546 (Allresist GmbH) for 1 min. Reaction ion etching (RIE) is performed to transfer the nanostructures to a-Si film. The residue resist is removed by immersing the chip first in acetone for 5~min, then in isopropanol for 5~min and finally in deionized water. 

\noindent\textbf{Experimental setup to implement SLST with metasurface. }Metasurface is fixed on a piece of hollow plastic, which can be adjusted in six degrees of freedom through a six-dimensional rotation stage. Objective lens with $20 \times$ magnifying factor and tube lens with the focal length of 200 mm is used as a microscope, enlarging the distance of six spots focused by metasurface from 70~$\mu$m to 1.9~mm. Then, three prisms at different heights are applied to separate six light beams. Four mini lenses with $f=15$~mm and two mini lenses with $f=30$~mm are used to couple the six beams into six multi-mode fibers with the core diameter of $62.5~\mu$m.

\section*{Data Availability.}
The data generated in this study have been deposited in the Zenodo database with the identifier \href{https://zenodo.org/records/10674374}{[DOI 10.5281/zenodo.10674373]}.

\section*{Code availability.}
The codes used for data analysis and simulation in this study have been deposited in the Zenodo database with the identifier \href{https://zenodo.org/records/10674374}{[DOI 10.5281/zenodo.10674373]}.

\section*{Acknowledgements.}
The authors thank Xiaoqi Zhou for insightful discussion. The authors thank the anonymous reviewers for the insightful comments on the work. K.~A., T.~Z. and H.~L. were supported by the National Key Research and Development Program of China~(Grant No.~2019YFA0308200), the National Natural Science Foundation of China~(Grants Nos.~11974213 and 92065112), Shandong Provincial Natural Science Foundation~(Grant Nos.~ZR2020JQ05 and ZR2023LLZ005), Taishan Scholar of Shandong Province~(Grant No.~tsqn202103013), Shenzhen Fundamental Research Program~(Grants Nos.~JCYJ20190806155211142 and JCYJ20220530141013029), Shandong University Multidisciplinary Research and Innovation Team of Young Scholars~(Grant No.~2020QNQT) and Higher Education Discipline Innovation Project (`111')~(Grant No.~B13029). S.~L. and G.~W were supported by the National Natural Science Foundation of China~(Grants Nos.~62375282 and 62205370). Y.~Z. was supported by the National Natural Science Foundation of China~(Grants No.~12205048) and Innovation Program for Quantum Science and Technology~(Grant No.~2021ZD0302000).

\section*{Author contributions.}
H.~L. conceived and designed the experiment. K.~A. T.~Z. and H.~L. carried out the experiment of SLST. Z.~L., S.~L., L.~W., W.~Z. and G.~W designed and fabricated metasurface. Y.~Z., X.~Y. and H.~L. conducted the theoretical analysis. H.~L. and G.~W. supervised the project. All authors contributed to writing the manuscript.

\section*{Competing interests}
The authors have no competing interests.

\appendix
\onecolumngrid

\renewcommand{\figurename}{\textbf{Supplementary Fig}}
\renewcommand{\tablename}{\textbf{Supplementary Table}}
\renewcommand{\thetable}{\arabic{table}}
\renewcommand\thesection{\alph{section}}

\section*{Supplementary Note 1: Shadow tomography with POVM}
\subsection*{A. Constructing classical shadow with POVM}
Consider a $d$-level quantum system, a set of $L$ rank-one projectors $\{\ket{\psi_l}\bra{\psi_l}\in\mathbb{H}_d\}_{l=1}^{L}$ is called a quantum (complex projective) 2-design if the average value of second moment $(\ket{\psi_l}\bra{\psi_l})^{\otimes 2}$ over the set $\{\ket{\psi_l}\}$ is proportional to the projector onto the totally symmetric subspace of two copies
\begin{equation}\label{Eq:2design}
\frac{1}{L}\sum_{l=1}^{L}(\ket{\psi_l}\bra{\psi_l})^{\otimes 2}=\binom{d+1}{2}^{-1}P_{\text{Sym}^{(2)}},
\end{equation}
where $P_{\text{Sym}^{(2)}}=\frac{1}{2}(\mathds 1_d\otimes\mathds1_d+\mathbb{F})$ is the  projector of the symmetric subspace, and $\mathbb F$ is the swap operator acting on 2-copy as $\mathbb F\ket{v}\otimes\ket{w}=\ket{w}\otimes\ket{v}$ for all $\ket{v},\ket{w}\in\mathbb C^d$. 

Following the defining property~Supplementary Eq.~\eqref{Eq:2design}, each quantum 2-design is proportional to a POVM 
\begin{equation}\label{Eq:2designPOVM}
\mathbf E=\{E_l=\frac{d}{L}\ket{\psi_l}\bra{\psi_l}\}_{l=1}^{L},
\end{equation}
since the elements $E_l$ are positive semidefinite and satisfy $\sum_{l=1}^{L}E_l=\mathds1_d$. Measuring a quantum state $\rho$ results in one of the $L$ outcomes indexed by $l\in[L]$, and by Born's rule, the corresponding probability 
\begin{equation}
p_l=\pr(l|\rho)=\tr(E_l\rho).
\end{equation}

Hereafter we focus on the case of $d=2$. The POVM defined in Supplementary Eq.~\eqref{Eq:2designPOVM} (together with the preparation of the corresponding state $\ket{\psi_l}$) can be viewed as a linear map $\mathcal M: \mathbb{H}_2\to\mathbb{H}_2$ as follows
\begin{equation}\label{Eq:CSmap}
\begin{split}
\mathcal M(\rho)&=\sum_{l=1}^{L}\pr(l|\rho)\ket{\psi_l}\bra{\psi_l}\\
&=\frac{2}{L}\sum_{l=1}^{L}\bra{\psi_l}\rho\ket{\psi_l}\ket{\psi_l}\bra{\psi_l}\\
&=\frac{2}{L}\tr_1\left(\left(\sum_{l=1}^{L}(\ket{\psi_l}\bra{\psi_l})^{\otimes 2}\right)\mathds1_2\otimes \rho\right)\\
&=2\tr_1\left(\frac{1}{6}(\mathds 1_2\otimes\mathds1_2+\mathbb{F})\mathds1_2\otimes\rho\right)\\
&=\frac{1}{3}\tr_1\left(\mathds 1_2\otimes\rho+\rho\otimes\mathds1_2\right)\\
&=\frac{1}{3}\left(\rho+\tr(\rho)\mathds1_2\right).
\end{split}
\end{equation}
The inverse of this map is 
\begin{equation}\label{Eq:inversemap}
\mathcal M^{-1}(X)=3X-\tr(X)\mathds 1_2\  \forall X\in\mathbb H_2.
\end{equation} 

For a single experimental run by performing POVM on $\rho$, we obtain the random outcome $l$ with probability $\pr(l|\rho)$, and the classical shadow is constructed according to Supplementary Eq.~\eqref{Eq:inversemap} as
\begin{equation}\label{Eq:CSsingle}
\hat{\rho_l}=\mathcal M^{-1}(\ket{\psi_l}\bra{\psi_l})=3\ket{\psi_l}\bra{\psi_l}-\mathds 1_2.
\end{equation}

It exactly reconstructs the underlying quantum state $\rho$ in expectation
\begin{equation}
\begin{split}
\mathbb E(\hat{\rho_l})&=\sum_l\pr(l|\rho)(3\ket{\psi_l}\bra{\psi_l}-\mathds 1_2)\\
&=3\sum_l\pr(l|\rho)\ket{\psi_l}\bra{\psi_l}-\mathds1_2\\
&=\rho+\mathds 1_2-\mathds1_2\\
&=\rho,
\end{split}
\end{equation}
where the third line is due to Supplementary Eq.~\eqref{Eq:CSmap}.

For an $N$-qubit state $\rho\in\mathbb H_2^{\otimes N}$, the POVM $\mathbf E^{\otimes N}$ acts on each qubit independently yields outcome as a string $\bm l=l_1l_2 \cdots l_N$ with probability 
\begin{equation}
\begin{split}
\pr(\bm l|\rho)&=\tr\left(\rho \bigotimes_{n=1}^NE_{l_n}\right)\\
&=\left(\frac{2}{L}\right)^N\bra{\psi_{\bm l}}\rho\ket{\psi_{\bm l}},
\end{split}
\end{equation}
where $\ket{\psi_{\bm l}}=\ket{\psi_{l_1}}\otimes\ket{\psi_{l_2}}\otimes\cdots\otimes\ket{\psi_{l_N}}$. For a single experimental run, the classical shadow shows 
\begin{equation}
\begin{split}
\hat{\rho}&=\bigotimes_{n=1}^N\mathcal M^{-1} (\ket{\psi_{\bm l}}\bra{\psi_{\bm l}})\\
&=\bigotimes_{n=1}^N\mathcal M^{-1}(\ket{\psi_{l_n}}\bra{\psi_{l_n}})\\
&=\bigotimes_{n=1}^N\left(3\ket{\psi_{l_n}}\bra{\psi_{l_n}}-\mathds 1_2\right),
\end{split}
\end{equation}  
which is in a tensor-product form.

\subsection*{B. Calibration of quantum channel $\mathcal M$}
In practice, there are unavoidable noises in unitary operations and measurements. To address this issue, robust classical shadow~(RShadow) protocol was proposed to mitigate the noise in shadow tomography~\cite{Chen2021PRXQuantum}.  It is convenient to represent $\mathcal M$ as a matrix $\mathcal L_{\mathcal M}$ in Pauli-Liouville representation. Here a linear operator $X$ is represented by a column vector $\ket{X}\!\rangle_j=\tr(\sigma_jX)$ in the Pauli-basis with $\sigma_0=\mathds 1_2/\sqrt{2}$ and $\sigma_1$, $\sigma_2$, $\sigma_3$ being the Pauli matrix $X/\sqrt{2}$, $Y/\sqrt{2}$, $Z/\sqrt{2}$. In this way~Supplementary Eq.~\eqref{Eq:CSmap} can be expressed as
\begin{equation}
\mathcal L_{\mathcal M}\liou{\rho}=\sum_l\braketliou{E_l}{\rho}\liou{\psi_l}
=\frac{2}{L}\sum_l\liou{\psi_l}\braliou{\psi_l}\liou{\rho}.
\end{equation}

The matrix form $\mathcal L_{\mathcal M}$ of the channel $\mathcal M$ corresponds to the projector onto the subspace spanned by $\mathbf{E}$, which is given by
\begin{equation}
\mathcal L_\mathcal{M}=\frac{2}{L}\sum_{l=1}^{L}\liou{\psi_l}\langle\!\bra{\psi_l},
\end{equation}\label{Eq:LiouChannel}
and the classical shadow is 
\begin{equation}\label{Eq:CSLiouChannel}
\liou{\hat{\rho}}=\mathcal L_\mathcal{M}^{-1}\liou{\psi_l},
\end{equation}
where $\mathcal L_\mathcal{M}^{-1}$ is the Moore-Penrose pseudo inverse of $\mathcal L_\mathcal{M}$~\cite{penrose1956}. 

The POVM $\mathbf{E}_{\text{octa}}=\{\frac{1}{3}\ket{\psi_l}\bra{\psi_l}: l=1, \cdots, 6\}$ with corresponding normalized vector $\ket{\psi_l}\in\{\ket{H}, \ket{V}, \ket{+}, \ket{-}, \ket{R}, \ket{L}\}$. The matrix $\mathcal L_{\mathcal M}$ and its inverse $\mathcal L_{\mathcal M}^{-1}$ are
\begin{equation}\label{app:MLM}
\mathcal L_{\mathcal M}=\begin{pmatrix}
1 & 0 & 0 & 0 \\
0 & \frac{1}{3} & 0 & 0 \\
0 & 0 & \frac{1}{3} & 0 \\
0 & 0 & 0 & \frac{1}{3} 
\end{pmatrix},\ \ 
\mathcal L_{\mathcal M}^{-1}=\begin{pmatrix}
1 & 0 & 0 & 0 \\
0 & 3 & 0 & 0 \\
0 & 0 & 3 & 0 \\
0 & 0 & 0 & 3 
\end{pmatrix}.
\end{equation}  

Accordingly, one has
\begin{equation}
\mathcal L_\mathcal{M}\liou{\rho}=\frac{1}{3}(\liou{\rho}+\liou{\mathds 1_2}),
\end{equation}
which is a depolarizing channel. Accordingly, the classical shadow in Supplementary Eq.~\eqref{Eq:CSsingle} can be written as
\begin{equation}
\begin{split}
\liou{\hat{\rho}}&=\mathcal L_\mathcal{M}^{-1}\liou{\psi_l}=3\liou{\psi_l}-\liou{\mathds 1_2}\Leftrightarrow\hat{\rho}=3\ket{\psi_l}\bra{\psi_l}-\mathds1 _2.
\end{split}
\end{equation}

Similarly, for an N-qubit state $\rho$,
\begin{equation}
\begin{split}
\mathcal L_{\mathcal M}\liou{\rho}&=\sum_{\bm l}\pr(\bm l|\rho)\liou{\psi_{\bm l}}\\
&=\sum_{l_1,l_2, \cdots, l_N}\braketliou{\bigotimes_{n=1}^NE_{l_n}}{\rho}\liou{\bigotimes_{n=1}^N\psi_{l_n}}\\
&=\bigotimes_{n=1}^N\left[\frac{1}{3}\sum_{l_n}\liou{\psi_{l_n}}\braliou{\psi_{l_n}}\right]\liou{\rho}\\
&=\bigotimes_{n=1}^N\mathcal L_{\mathcal M_{n}}\liou{\rho}.
\end{split}
\end{equation}

Then we have 
\begin{equation}\label{Eq:Qchannelmulti}
\mathcal L_{\mathcal M}=\bigotimes_{n=1}^N\mathcal L_{\mathcal M_{n}},
\end{equation}
and the classical shadow is 
\begin{equation}
\begin{split}
\liou{\hat{\rho}}&=\mathcal L_{\mathcal M}^{-1}\bigotimes_{n=1}^N\liou{\psi_{l_n}}\\
&=\bigotimes_{n=1}^N\mathcal L_{\mathcal M_{n}}^{-1}\liou{\psi_{l_n}}\\
&=\bigotimes_{n=1}^N(3\liou{\psi_{l_n}}-\liou{\mathds 1_2}).
\end{split}
\end{equation}

The quantum channel in Supplementary Eq.~\eqref{Eq:Qchannelmulti} can be written as 
\begin{equation}
\mathcal L_{\mathcal M}=\sum_{\lambda\in \{ 0,1 \}^N }f_{\lambda}\Pi_{\lambda},
\end{equation}
where $\lambda$ is an $n$-bit vector denoting the subspaces due to the irreducible representation, and $\Pi_{\lambda}=\bigotimes_{n=1}^{N}\Pi_{\lambda_n}$ in the tensor-product form with 
\begin{equation}
\Pi_{\lambda_n} = 
\begin{cases}
\liou{ \sigma_0}\braliou{\sigma_0},  & \lambda_n=0 \\
\mathds{1}_4 - \liou{\sigma_0}\braliou{\sigma_0}, & \lambda_n=1.
\end{cases}
\end{equation}

Here $\sigma_0=\mathds{1}_2/\sqrt{2}$ is the normalized single-qubit identity operator such that $\braliou{\sigma_0}\sigma_0\rangle\!\rangle =1$, and $\mathds{1}_4$ is the 4-dimensional identity matrix for this single-qubit operator space. That is, $\Pi_{0}$ subspace corresponds to the identity operator, and $\Pi_{1}$ is the complementary subspace spanned by the Pauli operators evenly. In the noiseless case, 
\begin{equation}
f_{\lambda}=\frac{1}{3^{|{\lambda}|}},
\end{equation}
where $|{\lambda}|$ is the number of $1$s in the $n$-bit vector $\lambda$. For the single-qubit case with $N=1$, the matrix form of $\mathcal L_{\mathcal M}$ is just shown in Supplementary Eq.~\eqref{app:MLM}.

Considering the noise (or imperfections) in practice, suppose the corresponding quantum channel of the noisy POVM can be written as
$\mathcal L_{\widetilde{\mathcal M}}=\sum_\lambda \tilde{f}_\lambda\Pi_\lambda$, which is still diagonal according to the subspaces. The noisy parameters $\tilde{f}_\lambda$ can be experimentally calibrated. 

To this end, a high-fidelity $N$-qubit product state 
\begin{equation}
\rho_{\mathbf 0}=\bigotimes_{n=1}^N\ket{0}\bra{0}\Leftrightarrow\liou{\rho_{\mathbf 0}}=\bigotimes_{n=1}^N\frac{1}{2}(\liou{\mathds 1_2}+\liou{Z}),
\end{equation}
is prepared and measured with noisy POVM, and the probability is 
$\tilde{\pr}({\bm l})=\braketliou{\tilde{E}_{\bm l}}{\rho_{\mathbf 0}}$ with $\liou{\tilde{E}_{\bm l}}$ being the noisy POVM.

Here we define $\liou{P_\lambda}=\bigotimes_{n=1}^N \liou{Z^{\lambda_n}}$, we can use $\braketliou{P_\lambda}{\psi_{\bm l}}$ to estimate the noisy parameter $\tilde{f}_\lambda$. To show this is indeed an unbiased estimator, we take the expectation value and find that
\begin{equation}
\begin{split}
\mathbb E \braketliou{P_\lambda}{\psi_{\bm l}}=&\sum_{\bm l}\braketliou{P_\lambda}{\psi_{\bm l}}\braketliou{\tilde{E}_{\bm l}}{\rho_{\mathbf 0}}\\
=&\braliou{P_\lambda} \mathcal L_{\widetilde{\mathcal M}} \liou{\rho_{\mathbf 0}}\\
=&\frac{1}{2^N}\left(\bigotimes_{n=1}^N \braliou{Z^{\lambda_n}}\right) \sum_{\lambda'} \tilde{f}_{\lambda'} \Pi_{\lambda'} \bigotimes_{n=1}^N (\liou{\mathds 1_2}+\liou{Z})\\
=&\frac{1}{2^N}\left(\bigotimes_{n=1}^N \braliou{Z^{\lambda_n}}\right) \sum_{\lambda'} \tilde{f}_{\lambda'} \bigotimes_{n=1}^N \liou{Z^{\lambda'_n}}=\tilde{f}_{\lambda}.
\end{split}
\end{equation}

Experimentally, we record the results of $\bm l$ from the noisy POVM, and then calculate the value $\braketliou{P_\lambda}{\psi_{\bm l}}$ on the classical computer which serves as an estimation of $\tilde{f}_{\lambda}$. One can repeat this process $M^\prime$ times, and average them to make the estimation more accurate. 
In this way, we can get the noisy parameters $\tilde{f}_{\lambda}$ of the noisy channel $\widetilde{\mathcal M}$, and the matrix form of its inverse $\widetilde{\mathcal M}^{-1}$ is
\begin{equation}
\mathcal L_{\mathcal M}^{-1}=\sum_{\lambda\in \{ 0,1 \}^N }\tilde{f}^{-1}_{\lambda}\Pi_{\lambda}.
\end{equation} 
In post-processing, we can use this new inverse channel $\widetilde{\mathcal M}$ instead of $\mathcal M$, such that the error in the noisy POVM could be mitigated.

\subsection*{C. Gate-dependent noise}
In Ref.~\cite{Chen2021PRXQuantum}, two main assumptions are made on the noise of measurement device to make the theoretical framework rigorous and analytical, i.e.,  
\begin{itemize}
    \item [\textbf{A1.}] The noise in the circuit is gate-independent, time-stationary, Markovian noise.
    \item [\textbf{A2.}] The experimental device can generate the computational basis state $\ket{\bm{0}}\equiv\ket{0}^{\otimes N}$ with sufficiently high fidelity. 
\end{itemize} 
Our experimental device is able to generate $\ket{HH}$ with high fidelity~($>0.99$) so that \textbf{A2} is satisfied. Metasurface is a passive optical device so that the metasurface-induced noise is time-stationary and Markovian noise as well. Note that measurement errors induced by metasurface are not strictly gate-independent, as the errors in $\sigma_x$, $\sigma_y$, and $\sigma_z$ measurements are $0.086\pm 0.005$, $0.073\pm 0.005$, and $0.101\pm 0.005$, respectively. However, it has been shown robust shadow tomography still works for gate-dependent noise~\cite{Chen2021PRXQuantum}. Our experimental results~(Figure 4 in main text) confirm this claim as well. To further confirm this point, we simulate the SLST and robust SLST on single-qubit state
\begin{equation}\label{Eq:noisystate}
    \rho^\prime=\left(1-\frac{\delta_x+\delta_y+\delta_z}{3}\right)\rho+\frac{\delta_x}{3}\sigma_x\rho\sigma_x+\frac{\delta_y}{3}\sigma_y\rho\sigma_y+\frac{\delta_z}{3}\sigma_z\rho\sigma_z.
\end{equation}
In each single run in simulation, the value of $\delta_{x,y,z}$ is randomly resampled from Gaussian distribution with mean value of $\bar{\delta}$ and standard deviation of $\sigma$. The simulation is equivalent to performing measurement with gate-dependent noises on the ideal state $\rho$. As shown in Supplementary Fig.~\ref{fig:fidelity_vs_sigma}~(a), the enhancement of robust SLST is not obvious when the noise is weak~($\bar{\delta}=0.05$). Robust SLST exhibits advantage when the noise is strong~($\bar{\delta}=0.1$) as shown in Supplementary Fig.~\ref{fig:fidelity_vs_sigma}~(b). The simulation results agree well with the experimental results shown in Figure 4 in main text, where the robust SLST significantly enhances the accuracy when optical loss is high.

\begin{figure}[h!]
\centering
\includegraphics[width=1.0\textwidth]{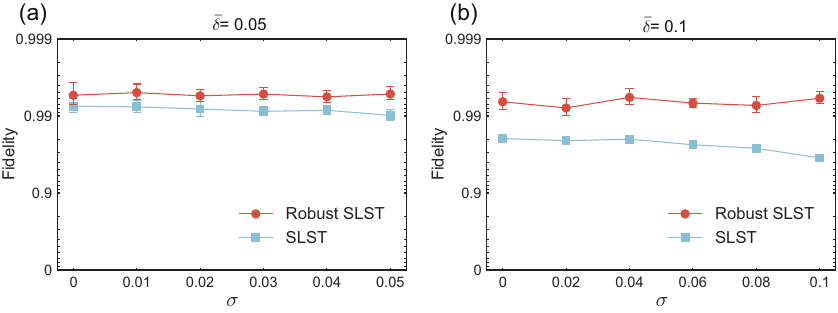}
\caption{\textbf{The simulation of robust SLST and SLST on the state in~Supplementary Eq.~\eqref{Eq:noisystate}.} (a), The simulation is performed by setting $\bar{\delta}=0.05$ and $\sigma\in[0,0.05]$ with interval of 0.01. (b), The simulation is performed by setting $\bar{\delta}=0.1$ and $\sigma\in[0,0.1]$ with interval of 0.02. The simulation is carried out with $M=2000$ runs. In robust SLST, additional $M^\prime=2000$ runs are used for calibration. We set $k=100$, hyperparameters $a_1 = 34.1$ and $b_1 = 5.7$. Error bars represent standard deviations obtained by repeating the simulation 5 times.}
\label{fig:fidelity_vs_sigma}
\end{figure}

\subsection*{D. Shadow norm}
For a given POVM $\mathbf E$, the shadow norm of observable $O$ is derived from the variance of the estimation $\hat{o}$ on state $\rho$. The variance Var($\hat{o}$) can be ideally written as
\begin{equation}\label{Seq:variance}
\text{Var}(\hat{o})=\sum_{l=1}^{L} \tr(\hat{\rho}_l O)^2 \tr(\rho E_l)-\tr(\rho O)^2.
\end{equation}

The shadow norm is then defined by the maximization of variance $\text{Var}(\hat{o})$ over $\rho$
\begin{equation}\label{Seq:variance_max}
\text{Var}(\hat{o})=\sum_{l=1}^{L} \tr(\hat{\rho}_l O)^2 \tr(\rho E_l)-\tr(\rho O)^2 \leq \max_{\rho}\sum_{l=1}^{L} \tr(\hat{\rho}_l O)^2 \tr(\rho E_l)-\tr(\rho O)^2.
\end{equation}

Note that the second term $\tr(\rho O)^2$ is a constant and can be ignored. Then, the shadow norm of $O$ is calculated by
\begin{equation}\label{Eq:shadownorm}
||O||^2_\text{shd}=\lambda_\text{max}\left\{\sum_{l=1}^{L}\tr(\hat{\rho}_l O)^2 E_l\right\},
\end{equation}
with $\lambda_\text{max}\left\{\cdot\right\}$ being the maximal eigenvalue of corresponding operator. 
In theoretical investigations such as~\cite{Nguyen2022PRL}, it is convenient to calculate shadow norm in Supplementary Eq.~\eqref{Eq:shadownorm}. For octahedron POVM, the theoretical shadow norm calculated according to Supplementary Eq.~\eqref{Eq:shadownorm} is $||O||^2_\text{shd}=1.5$.

Experimentally, the variance of estimator $\hat{o}$ is observed by  
\begin{equation}\label{Eq:varexp}
\text{Var}(\hat{o}) = \frac{1}{M}\sum_{m=1}^{M} \left(\hat{o}^{(m)} - \hat{O} \right)^2.  
\end{equation}  

Without consideration of experimental noise, Supplementary Eq.~\eqref{Eq:varexp} converges to Supplementary Eq.~\eqref{Seq:variance} when $M\to\infty$. For octahedron POVM, the maximization of Supplementary Eq.~\eqref{Seq:variance} over single-qubit pure state $\rho$ yields  $||O||^2_\text{shd}=0.75$, which is considered as the ideal value for experimentally observed maximal variance in Supplementary Eq.~\eqref{Eq:varexp}.

\subsection*{E. Simulation of shadow norm with SIC POVM} The SIC POVM in qubit system can be expressed by $\mathbf E_\text{SIC} = \{\frac{1}{2}\ket{\psi_l}\bra{\psi_l}\}_{l=1}^{4}$ with
\begin{equation}
\begin{aligned}
& \left|\psi_1\right\rangle=|0\rangle \\
& \left|\psi_2\right\rangle=\frac{1}{\sqrt{3}}|0\rangle+\sqrt{\frac{2}{3}}|1\rangle \\
& \left|\psi_3\right\rangle=\frac{1}{\sqrt{3}}|0\rangle+\sqrt{\frac{2}{3}} e^{i \frac{2 \pi}{3}}|1\rangle \\
& \left|\psi_4\right\rangle=\frac{1}{\sqrt{3}}|0\rangle+\sqrt{\frac{2}{3}} e^{i \frac{4 \pi}{3}}|1\rangle.
\end{aligned}
\end{equation}

For fixed observable $O=\ket{\psi_{\kappa,\nu}}\bra{\psi_{\kappa,\nu}}\in\mathbf O$, single-qubit state $\rho=\ket{\psi_{\gamma,\phi}}\bra{\psi_{\gamma,\phi}}$ and $\mathbf{E}_\text{SIC}$, we simulate the statistic of outcomes with $M$ runs, and calculate the variance according to Supplementary Eq.~\eqref{Eq:varexp}. Then, the shadow norm is calculated by maximization over state set $\mathbf{P}$, i.e., $\max_\mathbf{P}\text{Var}(\hat{o}^{(m)})$, where $\mathbf{P}$ is the set of 20 pure states.

\section*{Supplementary Note 2: Details of SLST}
\subsection*{A. Loss function in SLST}
Generally, the loss function in SLST is squared Frobenius norm between two matrices $\tau$ and $\rho$ defined  as 
\begin{equation}
    \begin{aligned}
      N_F(\tau,\rho)=\|\rho-\tau\|^2_F=\tr(\rho-\tau)^2=\tr(\rho^2)+\tr(\tau^2)-2\tr(\rho\tau).
    \end{aligned}
\end{equation}

For this loss function, we can write down the unbiased estimator with shadows $\{\hat{\rho}^{(m)}\}_{m=1}^{M}$ as follows. For the first term, it shows 
\begin{equation}
    \begin{aligned}
       \frac{2}{M(M-1)}\sum_{m<n}\tr\left[\hat{\rho}^{(m)}\hat{\rho}^{(n)}\right], 
    \end{aligned}
\end{equation}
which is an order-2 polynomial function of $\hat{
\rho}$~\cite{Huang2020NPhysics}. Obviously, $\tr(\hat{\rho}\tau)$ is an unbiased estimator as well. Then, the unbiased estimator of $N_F$ in total is 
\begin{equation}
    \begin{aligned}
       \hat{N}_F(\tau)=\frac{2}{M(M-1)}\sum_{m<n}\tr\left[\hat{\rho}^{(m)}\hat{\rho}^{(n)}\right]+ \tr(\tau^2)-2 \sum_{m}\tr(\hat{\rho}^{(m)}\tau). 
    \end{aligned}
\end{equation}

Accordingly, the gradient is
\begin{equation}\label{Eq:gradientN}
\bm{g}_k = \frac{\hat{N}_F(\bm{r}_k+B_k\boldsymbol{\Delta}_k) - \hat{N}_F(\bm{r}_k-B_k\boldsymbol{\Delta}_k)}{2 B_k}\boldsymbol{\Delta}_k.
\end{equation}

Note that fidelity $F$ can also be used as loss function if $\tau$ is pure state, as $\hat{F}(\tau)$ is an unbiased estimation with classical shadows $\{\hat{\rho}^{(m)}\}$
\begin{equation}
\hat{F}(\tau)=\sum_m\tr\left(\hat{\rho}^{(m)}\tau\right),
\end{equation}
and the gradient is 
\begin{equation}\label{Eq:gradientF}
\bm{g}_k = \frac{\hat{F}(\bm{r}_k+B_k\boldsymbol{\Delta}_k) - \hat{F}(\bm{r}_k-B_k\boldsymbol{\Delta}_k)}{2 B_k}\boldsymbol{\Delta}_k.
\end{equation} 
\subsection*{B. Setting of hyperparameters in SPSA optimization}
The setting of hyperparameters $a_1$, $a_2$, $a_3$, $b_1$ and $b_2$ determines the convergence of SPSA optimization. Previous investigations have concluded that $a_3=0.602, b_2=0.101$ are generally good choices for most optimization tasks~\cite{Spall1992,Spall1998,Maryak2001}, so that we set $a_3=0.602, b_2=0.101$ in our optimization. Besides, we find that $a_2$ is trivial compared with other four hyperparameters so that we set $a_2=0$.  $a_1$ and $b_1$ are determined through numerical simulations. We set $a_1=13$, $b_1=0.5$ in single-photon experiment, while $a_1=8.5$ and $b_1=1.4$ in two-photon experiment.

\subsection*{C. Model of proposed state $\tau_k$ in SLST}
For proposed state $\tau_k$ being an $N$-qubit pure state $\tau_k=\ket{\zeta}\bra{\zeta}$ with dimension $d=2^N$, it can be formalized by 
\begin{equation}\label{Eq:purestate}
\ket{\zeta} =\frac{1}{\sqrt{\sum_{i=1}^dr_i^2}} \left(
\begin{array}{cc}
    r_1 \\
    r_2 e^{ir_{d+1}} \\
    r_3 e^{ir_{{d+2}}} \\
    \vdots \\
    r_{d} e^{ir_{2d-1}} \\
\end{array}
\right),
\end{equation}
and we set $\bm{r}_k= \{r_1, r_2, \cdots, r_{2d-1}\}$ with $r_i\in\mathbb{R}$ in SPSA optimization. 

For general case, a mixed state, the proposed state $\tau_k$ is modeled by Cholesky decomposition
\begin{equation}\label{Eq:Cholesky}
\tau_k = \frac{T T^\dagger }{\tr(T T^\dagger)},
\end{equation}
where $T$ is a lower triangular matrix
\begin{equation}\label{supeq6}
T =\begin{pmatrix}
    r_1 & 0   & \cdots & 0 \\
    r_{d+1}+i r_{d+2} & r_2 & \cdots & 0 \\
    \vdots & \vdots & \ddots & \vdots \\
    r_{d^2-1}+i r_{d^2} & r_{d^2-3}+i r_{d^2-2} & \cdots & r_{d} \\
\end{pmatrix}.
\end{equation}

Accordingly, we set $\bm{r}_k = \{ r_1, r_2, \cdots, r_{d^2} \}, r_i \in \mathbb{R}$ in SPSA optimization.

\subsection*{D. Scaling of SLST}
To investigate the scaling of SLST, we simulate SLST with POVM $\mathbf E_{\text{octa}}$ on 50 randomly generated $N$-qubit pure states $\rho_N$ with $N=2, 4, 6$ and $8$, respectively. The results of infidelity $1-F(\tau_k,\rho_N)$ are shown in Supplementary Fig.~\ref{Fig:simulation_multiqubit}, in which we set the iterations $k=200, 1000, 5000$ and $25000$ for $N=2$, $4$, $6$ and $8$, respectively. The extracted scaling of SLST is $O(d\log d/M)$, which is slightly worse than $O(d/M)$ in SQT and $O(d^\eta/M) (\eta>1)$ in SGQT. However, SLST with POVM requires only one experimental setting and the POVM is locally implemented on individual qubit, which is friendly to experiment. A comparison of these technologies is shown in Table~\ref{Tab:complexity}. 
\begin{figure}[h!tbp]%
\centering
\includegraphics[width=0.6\linewidth]{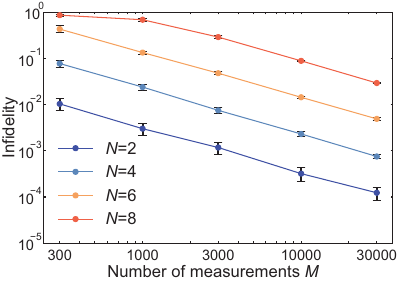}
\caption{\textbf{The average infidelity of 50 reconstructed states with number of measurements $M$.} The hyperparameters we set in SPSA optimization are $a_1=20, b_1=0.35$ for $N=2$, $a_1=15,b_1=0.79$ for $N=4$, $a_1=30,b_1=0.92$ for $N=6$ and $a_1=77, b_1=0.92$ for $N=8$. The error bars are the standard deviations of infidelity over 50 states.}
\label{Fig:simulation_multiqubit}
\end{figure}
\begin{table*}[ht!bp]
	\centering
	\begin{tabular}{c| c| c| c| c}\hline\hline
		 & Infidelity & Experimental Setting & Measurement Type & Online/Offline\\\hline
        SQT & $O(d/M)$ & $O(d)$ & Global projective measurement& Offline\\\hline
        SGQT & $O(d^{\eta}/M)$ & $O(M)$ & Global projective measurement& Online\\\hline
        SLST with POVM& $O(d \log d/M)$ & $1$ & Local POVM& Online\\\hline\hline
	\end{tabular}
	\caption{Comparison of SQT, SGQT and SLST with POVM.}
\label{Tab:complexity}
\end{table*}

\subsection*{E. Initial setting of $\tau_0$}
Generally, achieving the global minimum instead of local minimum is challenging in optimization. Indeed, SPSA optimization avoids local minimum under asymptotic iterations due to stochastic perturbation~\cite{Maryak2001}. However, SPSA optimization does not guarantee the global convergence in each iteration~\cite{Spall1998, Maryak2001}. We show that the global convergence in each iteration can be improved by setting initial $\tau_0$ with prior information, instead of randomly setting initial $\tau_0$.

The direct estimation from classical shadows $\hat{\rho}=\frac{1}{M}\sum_{m=1}^{M}\hat{\rho}^{(m)}$ returns a Hermitian matrix, which has an eigen-decomposition in form of   
\begin{equation}
\hat{\rho}=\sum_{i=1}^d \lambda_i\ket{\Psi_i}\bra{\Psi_i},
\end{equation} 
with $\lambda_i$ and $\ket{\Psi_i}$ being the eigenvalue and eigenvector of $\hat{\rho}$. However, $\hat{\rho}$ is not a semi-positive matrix so that $\lambda_i$ might be negative. We set the initial $\tau_0$ as 
\begin{equation}
\tau_0=\frac{\sum_{i=1}^d |\lambda_i|\ket{\Psi_i}\bra{\Psi_i}}{\sum_{i=1}^d|\lambda_i|},
\end{equation}
which is close to $\hat{\rho}$. Then, we do Cholesky decomposition on $\tau_0$ according to Supplementary Eq.~\eqref{Eq:Cholesky}, and obtain the corresponding $\bm{r}_0$ to start SPSA optimization. 

To show the advantage of this modification, we simulate the SLST with $M=2000, 3500$ and $5000$ runs on randomly generated 2-, 3- and 4-qubit mixed states, respectively. The results are shown in Supplementary Fig.~\ref{Fig:Inital_guess}. We observe that the initial setting of $\tau_0$ significantly influences efficiency and accuracy of SLST. With modified $\tau_0$, SLST converges more quickly and achieves lower infidelity than that with random setting of $\tau_0$.

\begin{figure*}[h!t]%
\centering
\includegraphics[width=\linewidth]{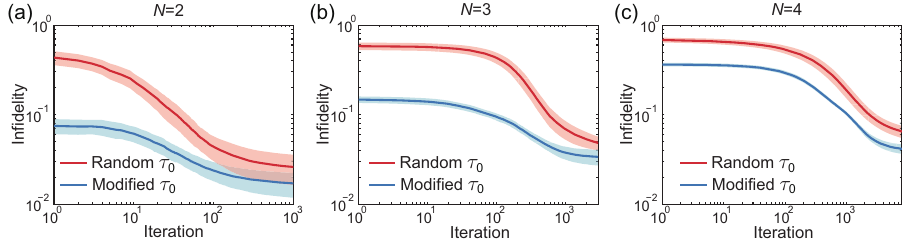}
\caption{\textbf{Simulation of SLST with different initial $\tau_0$.} Blue line and red line represent the infidelity against iteration with modified $\tau_0$ and random $\tau_0$ of SLST on \textbf{(a)} $N=2$, \textbf{(b)} $N=3$ and \textbf{(c)} $N=4$ mixed states, respectively. The hyperparameters we set in SPSA with modified $\tau_0$ are $a_1=4.3,b_1=0.5$ for $N=2$, $a_1=6.2,b_1=1.4$ for $N=3$ and $a_1=16.4, b_1=2.8$ for $N=4$. The hyperparameters we set in SPSA with random $\tau_0$ are $a_1=48, b_1=1.1$ for $N=2$, $a_1=44, b_1=8.0$ for $N=3$ and $a_1=49, b_1=12.7$ for $N=4$. The shadings represent the standard deviation of infidelities over 100 randomly generated mixed states.}
\label{Fig:Inital_guess}
\end{figure*}

\section*{Supplementary Note 3: Experimental demonstration of SGQT on single-photon state with metasurface}
Self-guided quantum tomography~(SGQT) is an iterative protocol to reconstruct underlying state of $\rho$. In fact, it is a SPSA optimization of the problem
\begin{equation}
\label{Eq:SGQTopt}
    \begin{split}
        \text{maximize}\hspace{1.2cm}&F(\tau^\text{SGQT})=\sqrt{\tr(\rho\tau^\text{SGQT})}\\
        \text{subject to}\hspace{1.2cm}&\tau^\text{SGQT}\geq0,\hspace{2pt}\tr(\tau^\text{SGQT})=1,
    \end{split}
\end{equation}
where $\tau^\text{SGQT}$ is the proposed state that is positive semidefinite ($\tau\geq0$) with unit trace ($\tr(\tau^\text{SGQT})=1$), and the loss function is the squared state fidelity $F(\tau^\text{SGQT})$. Note that $\tau^\text{SGQT}$ is restricted to pure states in SGQT. As an $N-$qubit pure state $\tau^\text{SGQT}$ can be modeled with $2d-1$ parameters, so that we denote $\tau^\text{SGQT}$ by a $2d$-dimensional vector $\bm{r}=[r_1, r_2,\cdots, r_{2d}]$. Accordingly, squared state fidelity is denoted by $F(\bm{r})=F(\rho,\tau)$, which equals to the probability of projection $\rho$ on $\tau^\text{SGQT}$.

The process of SGQT algorithm can be described as follows. 
\begin{enumerate}
\item Randomly guessing an initial state $\tau_0^\text{SGQT}$; 
\item Perturb $\tau_0(\bm{r})$ with a random perturbation vector $\boldsymbol{\Delta}_k =[\Delta_{k1}, \Delta_{k2}, \cdots, \Delta_{kd^2}]$ with $\Delta_{ki}$ being generated from Bernoulli $\pm1$ distribution with equal probability; 
\item Implement projective measurement on perturbed states $\tau(\bm{r}_k+B_k\boldsymbol{\Delta}_k)$ and $\tau(\bm{r}_k-B_k\boldsymbol{\Delta}_k)$;
\item Calculate the gradient by 
\begin{equation}\label{Eq:SGQTgradient}
\bm{g}_k = \frac{F(\bm{r}_k+B_k\boldsymbol{\Delta}_k) - F(\bm{r}_k-B_k\boldsymbol{\Delta}_k)}{2 B_k}\boldsymbol{\Delta}_k,
\end{equation}
and update $\bm{r}_{k}$ to $\bm{r}_{k+1} = \bm{r}_{k}+A_{k}\bm{g}_k$;
\item Repeat step 2-step 4 until $\bm{g}_k$ converges to zero, and corresponding $\tau_k$ is the reconstructed state with SGQT protocol.
\end{enumerate}

Experimentally, we demonstrate SGQT on single-photon state $\ket{\psi_{\gamma, \phi}}\in\mathbf{P}$ with an E-QWP, an E-HWP and metasurface. Note that only the photon passing through region 1 of the metasurface is post-selected in SGQT, which acts as a PBS. The projective measurement on $\tau^\text{SGQT}$ and its orthogonal state $\tau^\text{SGQT}_\perp$ is realized by setting the angles of  E-QWP and E-HWP. Consequently, the region 1 of metasurface deflects $\tau^\text{SGQT}$ and $\tau^\text{SGQT}_\perp$ into opposite directions. By collecting the counts at two directions, we can calculate the probability of projection on $\tau^\text{SGQT}$ and $\tau^\text{SGQT}_\perp$, respectively. In each iteration, seven experimental runs are carried out to project $\ket{\psi_{\gamma, \phi}}$ on $\tau_{k+}^\text{SGQT}=\bm{r}_k+B_k \bm{\Delta}_k$ and $\tau_{k-}^\text{SGQT}=\bm{r}_k-B_k \bm{\Delta}_k$, i.e., 
four experimental runs for $\tau_{k+}^\text{SGQT}=\bm{r}_k+B_k \bm{\Delta}_k$ and three experimental runs for $\tau_{k-}^\text{SGQT}=\bm{r}_k-B_k \bm{\Delta}_k$, respectively.

\section*{Supplementary Note 4: Design, fabrication and characterization of metasurface}

\subsection*{A. Design of metasurface to realize POVM $\mathbf E_\text{octa}$}
The POVM $\mathbf E_\text{octa}$ is equivalent to randomly selecting three Pauli observables and then performing projective measurement on its eigenstates. To this end, the metasurface is designed to consist of three regions with same size ($210\mu m\times70\mu m$), each of which corresponds to the projective measurement of observable $\sigma_j, j\in\{x, y, z\}$. Each region is further designed to spatially separate eigenstates $\ket{\psi_{\sigma_j}^+}$ and $\ket{\psi_{\sigma_j}^-}$ of $\sigma_j$, which is achieved by individual phase control of $\ket{\psi_{\sigma_j}^+}$ and $\ket{\psi_{\sigma_j}^-}$ when they pass through metasurface by
\begin{equation}\label{Eq:metatransform}
\Phi_{\sigma_j}^\pm(x,y) = - \frac{2 \pi}{\lambda} \left(\sqrt{(x-x_{\sigma_j,0}^\pm)^2+(y-y_{\sigma_j,0}^\pm)^2+f^2}-f \right).
\end{equation}
$\Phi_{\sigma_j}^\pm(x,y)$ represents phase configuration at the output of metasurface with input polarization $\ket{\psi_{\sigma_j}^+}$ and $\ket{\psi_{\sigma_j}^-}$, $(x^+_{\sigma_j,0}, y^+_{\sigma_j,0})$ and $(x^-_{\sigma_j,0}, y^-_{\sigma_j,0})$ are the positions of separated focal spots of $\ket{\psi_{\sigma_j}^+}$ and $\ket{\psi_{\sigma_j}^-}$ at focal plane and $f$ is the focal length. Our aim is to design a metasurface to realize phase configuration of $\Phi_j^\pm(x,y)$ calculated according to Supplementary Eq.~\eqref{Eq:metatransform} with fixed $(x^\pm_{\sigma_j,0}, y^\pm_{\sigma_j,0})$, $\lambda$ and $f$. The parameters we set to calculate $\Phi_{\sigma_j}^\pm(x,y)$ are shown in Table~\ref{Tb:parametersetting}.
 \begin{table}[h!]
\begin{center}
\begin{tabular}{ |c|c| } 
 \hline
~Parameters &  Values \\\hline 
$x$ in $\Phi_{\sigma_x}^\pm(x,y)$ & $[-105\mu\text{m}, 105\mu\text{m}]$ with interval of 0.5$\mu\text{m}$ \\ \hline
$y$ in $\Phi_{\sigma_x}^\pm(x,y)$ & $[-35\mu\text{m}, 35\mu\text{m}]$ with interval of 0.5$\mu\text{m}$\\ \hline
$(x^\pm_{\sigma_x,0}, y^\pm_{\sigma_x,0})$ & $(\pm35\mu\text{m},0\mu\text{m})$\\ \hline
$x$ in $\Phi_{\sigma_y}^\pm(x,y)$ & $[-105\mu\text{m}, 105\mu\text{m}]$ with interval of 0.5$\mu\text{m}$ \\ \hline
$y$ in $\Phi_{\sigma_y}^\pm(x,y)$ & $[-105\mu\text{m}, -35\mu\text{m}]$ with interval of 0.5$\mu\text{m}$\\ \hline
$(x^\pm_{\sigma_y,0}, y^\pm_{\sigma_y,0})$ & $(\pm35\mu\text{m},-70\mu\text{m})$\\ \hline
$x$ in $\Phi_{\sigma_z}^\pm(x,y)$ & $[-105\mu\text{m}, 105\mu\text{m}]$ with interval of 0.5$\mu\text{m}$ \\ \hline
$y$ in $\Phi_{\sigma_z}^\pm(x,y)$ & $[35\mu\text{m}, 105\mu\text{m}]$ with interval of 0.5$\mu\text{m}$\\ \hline
$(x^\pm_{\sigma_z,0}, y^\pm_{\sigma_z,0})$ & $(\pm35\mu\text{m},70\mu\text{m})$\\ \hline
$f$ & $150\mu\text{m}$\\ \hline
$\lambda$ & $810\text{nm}$\\
 \hline
\end{tabular}
\end{center}
\caption{The values of parameters set in calculation of phase configurations $\Phi_{\sigma_j}^\pm(x,y)$ in Supplementary Eq.~\eqref{Eq:metatransform}.}\label{Tb:parametersetting}
\end{table}
\begin{figure*}[h]%
\centering
\includegraphics[width=0.8\linewidth]{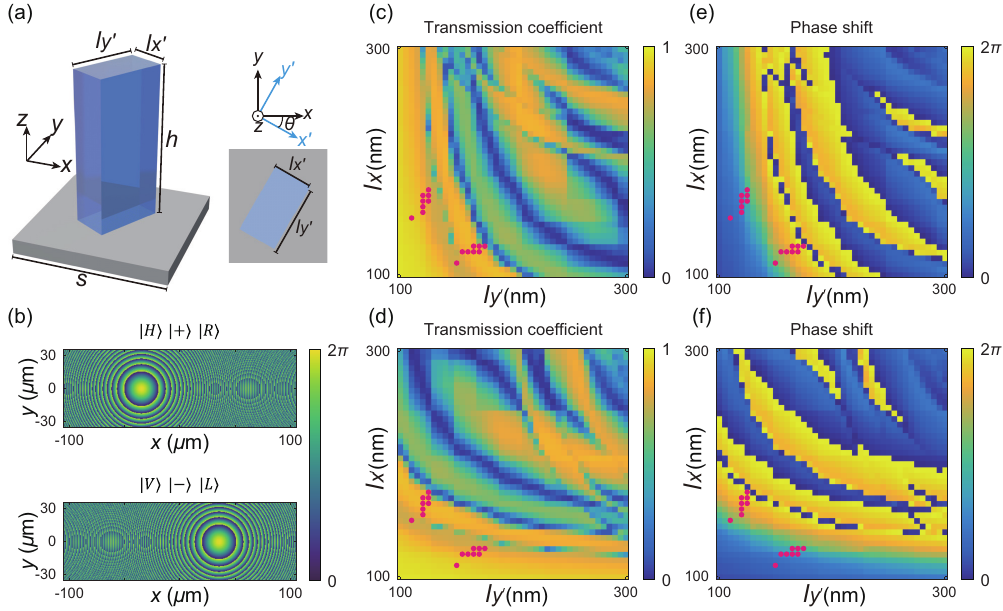}
\caption{\textbf{Schematic diagram and design of the metasurface.} \textbf{(a),} Schematic of the designed meta-atom consisting of an a-Si nanopillar on a fused-silica substrate. \textbf{(b),} Target phase configuration for orthogonally polarized states when light passes through metasurface. \textbf{(c)} and \textbf{(d),} Mapping of transmission coefficient for LP light polarized along $x^\prime$ and $y^\prime$, respectively, as a function of the parameters of $l_{x^\prime}$ and $l_{y^\prime}$ of the nanopillars. \textbf{(e)} and \textbf{(f),} Mapping of phase shift $\varphi_{x^\prime}$ and $\varphi_{y^\prime}$, respectively, as a function of the parameters of $l_{x^\prime}$ and $l_{y^\prime}$ of the nanopillars. }
\label{Fig:metadesign}
\end{figure*}

The calculated phase configurations of $\Phi_{\sigma_j}^{+}(x, y)$ and $\Phi_{\sigma_j}^{-}(x, y)$ are shown in Supplementary Fig.~\ref{Fig:metadesign}(b). To realize phase configurations $\Phi_{\sigma_j}^\pm(x, y)$, we design the metasurface consisting of $420\times140$ nanopillars with interval of $500$nm. As shown in Supplementary Fig.~\ref{Fig:metadesign}(a), the nanopillar located at position $(x_i, y_i)$ can be regarded as a polarization-dependent scatter acting on input polarization with transformation matrix 
\begin{equation}\label{Eq:metaunitary}
U_{\text{pillar}} =
\begin{pmatrix}
    \cos \theta & -\sin \theta \\
    \sin \theta & \cos \theta \\
\end{pmatrix}
\begin{pmatrix}
    e^{i\varphi_{x^\prime}} & 0 \\
    0 & e^{i\varphi_{y^\prime}} \\
\end{pmatrix}
\begin{pmatrix}
    \cos \theta & \sin \theta \\
    -\sin \theta & \cos \theta \\
\end{pmatrix}.
\end{equation}

Here, $\theta$ is the angle of nanopillar, $\varphi_{x^\prime}$ and $\varphi_{y^\prime}$ are the phase accumulated on the polarization component decomposed along the direction $x^\prime$ and $y^\prime$ respectively, which are determined by length $l_{x^\prime}$ and $l_{y^\prime}$. Thus, by properly designing the parameters $\{\theta, l_{x^\prime}, l_{y^\prime}\}$ of nanopillar at each $(x_i, y_i)$, we can realize the desired phase configuration $\Phi_{\sigma_j}^+(x,y)$ and $\Phi_{\sigma_j}^-(x,y)$ for polarization $\ket{\psi_{\sigma_j}^+}$ and $\ket{\psi_{\sigma_j}^-}$, respectively. 

The numerical simulation results of $\varphi_{x^\prime}$ and $\varphi_{y^\prime}$ are shown in Supplementary Fig.~\ref{Fig:metadesign}(e) and Supplementary Fig.~\ref{Fig:metadesign}(f) respectively, in which we set $l_{x^\prime}$ and $l_{y^\prime}$ from $100$nm to $300$nm with interval of $5$nm. The corresponding transmittances are shown in Supplementary Fig.~\ref{Fig:metadesign}(c) and Supplementary Fig.~\ref{Fig:metadesign}(d) respectively. The choice of $\{\theta, l_{x^\prime}, l_{y^\prime}\}$ depends on the polarization $\ket{\psi_{\sigma_j}^+}$ and $\ket{\psi_{\sigma_j}^-}$ we want to separate, and we will discuss separately.   

\subsubsection{Separation of linear polarizations}
For linear polarizations $\ket{\psi_{\sigma_z}^\pm}$ and $\ket{\psi_{\sigma_x}^\pm}$, we set $\theta$ of all nanopillars to be a constant. In this sense, phase configuration is determined by phase $\varphi_{x^\prime} $ and $\varphi_{y^\prime}$ in Supplementary Eq.~\eqref{Eq:metaunitary}. Specifically, to separate polarization $\ket{H}$ ($\ket{\psi_{\sigma_z}^+}$) and $\ket{V}$ ($\ket{\psi_{\sigma_z}^-}$), we set $\theta = 0^{\circ}$ that leads to $x^\prime=x$ and $y^\prime=y$. According to Supplementary Eq.~\eqref{Eq:metaunitary}, the transformation of $\ket{H}$ and $\ket{V}$ after a single nanopillar is
\begin{equation}\label{Eq:meta_HV}
U_{\text{pillar}}\ket{H} = e^{i \varphi_{x^\prime} }\ket{H}, U_{\text{pillar}}\ket{V} = e^{i \varphi_{y^\prime} }\ket{V}.
\end{equation}

According to the value of $\Phi_{\sigma_z}^\pm(x_i, y_i)$ at position $(x_i, y_i)$, we determine $l_{x^\prime}$ and $l_{y^\prime}$ by $\varphi_{x^\prime}\approx\Phi_{\sigma_z}^+(x_i, y_i)$ (Supplementary Fig.~\ref{Fig:metadesign}(e)) and $\varphi_{y^\prime}\approx\Phi_{\sigma_z}^-(x_i, y_i)$ (Supplementary Fig.~\ref{Fig:metadesign}(f)). Note that the choice of $(l_{x^\prime}, l_{y^\prime})$ is not unique, and the transmittances with corresponding $(l_{x^\prime}, l_{y^\prime})$ shown in Supplementary Fig.~\ref{Fig:metadesign}(c) and Supplementary Fig.~\ref{Fig:metadesign}(d) should be as high as possible. 

To separate polarizations $\ket{+}$ ($\ket{\psi_{\sigma_x}^+}$) and $\ket{-}$ ($\ket{\psi_{\sigma_x}^-}$), we set $\theta = 45^{\circ}$ for all nanopillars. According to Supplementary Eq.~\eqref{Eq:metaunitary}, the transformation of a single nanopillar is  
\begin{equation}\label{Eq:meta_+-}
U_{\text{pillar}}\ket{+} = e^{i \varphi_{x^\prime}}\ket{+}, U_{\text{pillar}}\ket{-} = e^{i \varphi_{y^\prime}}\ket{-}.
\end{equation}

Similar to the situation of $\Phi_{\sigma_z}^\pm(x_i, y_i)$, the value of $(l_{x^\prime}, l_{y^\prime})$ is determined by $\varphi_{x^\prime}\approx\Phi_{\sigma_x}^+(x_i, y_i)$ and $\varphi_{y^\prime}\approx\Phi_{\sigma_x}^-(x_i, y_i)$ along with the consideration of high transmittance.  

\subsubsection{Separation of circular polarizations}
The design to separate circular polarizations $\ket{L}$ ($\ket{\psi_{\sigma_y}^+}$) and $\ket{R}$ ($\ket{\psi_{\sigma_y}^-}$) is different with the situations of linear polarizations. To realize $\Phi_{\sigma_y}^\pm(x_i, y_i)$, $\varphi_{x^\prime}$ and $\varphi_{y^\prime}$ in Supplementary Eq.~\eqref{Eq:metaunitary} should satisfy $\lvert\varphi_{x^\prime} - \varphi_{y^\prime}\rvert = \pi$~\cite{li2019multidimensional}. With this constraint, a single nanopillar transforms $\ket{L}$ and $\ket{R}$ by 
\begin{equation}\label{Eq:meta_RL}
U_{\text{pillar}}\ket{L}=e^{i (\varphi_{x^\prime} + 2\theta)}\ket{R}, U_{\text{pillar}}\ket{R}=e^{i (\varphi_{x^\prime} - 2\theta)}\ket{L}.
\end{equation}

The values of $\theta$ and $\varphi_{x^\prime}$ are determined by 
\begin{equation}\label{Eq:meta_phirl2}
\left\{
\begin{aligned}
    \theta=\frac{\Phi_{\sigma_y}^+(x_i, y_i)- \Phi_{\sigma_y}^-(x_i, y_i)}{4} \\
    \varphi_{x^\prime}=\frac{\Phi_{\sigma_y}^+(x_i, y_i)+ \Phi_{\sigma_y}^-(x_i, y_i)}{2}\\
\end{aligned}
\right.
\end{equation}

For the convince of fabrication, we choose nanopillars with 16 different $(l_{x^\prime}, l_{y^\prime})$ shown with red dots in Supplementary Fig.~\ref{Fig:metadesign}(c)-(f), and the parameters are shown in Table~\ref{Tab:pillarsize}. For the phase $\Phi_{\sigma_y}^\pm(x_i, y_i)$ at position $(x_i, y_i)$, we calculate $\theta$ and $\varphi_{x^\prime}$ according to Supplementary Eq.~\eqref{Eq:meta_phirl2}, and choose the closest $\varphi_{x^\prime}$ in Table~\ref{Tab:pillarsize} for fabrication. 
\begin{table}[htbp]
	\centering
	\begin{tabular}{ c c c c c c}\hline\hline
		$l_{x^\prime} \ (\text{nm})$ & $l_{y^\prime} \ (\text{nm})$ & $\varphi_{x^\prime}\ / \ 2\pi$ & $\varphi_{y^\prime} \ / \ 2 \pi$ & $\mathbf{T}_{x^\prime}$ & $\mathbf{T}_{y^\prime}$\\\hline
        110 & 150 & 0.1544 & 0.5924 & 0.9365 & 0.7640\\
        \hline
        120 & 155 & 0.2594 & 0.7413 & 0.8770 & 0.8245\\
        \hline
        120 & 160 & 0.2718 & 0.7972 & 0.8710 & 0.8567\\
        \hline
        120 & 165 & 0.2948 & 0.8546 & 0.8646 & 0.8837\\
        \hline
        120 & 170 & 0.3108 & 0.9024 & 0.8579 & 0.8803\\
        \hline
        125 & 165 & 0.4453 & 0.8751 & 0.8087 & 0.8857\\
        \hline
        125 & 170 & 0.4723 & 0.9251 & 0.8016 & 0.8618\\
        \hline
        125 & 175 & 0.4963 & 0.9698 & 0.7950 & 0.7676\\
        \hline
        150 & 110 & 0.5924 & 0.1544 & 0.7640 & 0.9365\\
        \hline
        155 & 120 & 0.7413 & 0.2594 & 0.8245 & 0.8770\\
        \hline
        160 & 120 & 0.7972 & 0.2718 & 0.8567 & 0.8710\\
        \hline
        165 & 120 & 0.8546 & 0.2948 & 0.8837 & 0.8646\\
        \hline
        165 & 125 & 0.8751 & 0.4453 & 0.8857 & 0.8087\\
        \hline
        170 & 120 & 0.9024 & 0.3108 & 0.8803 & 0.8579\\
        \hline
        170 & 125 & 0.9251 & 0.4723 & 0.8618 & 0.8016\\
        \hline
        175 & 125 & 0.9698 & 0.4963 & 0.7676 & 0.7950\\
        \hline
	\end{tabular}
	\caption{Selected nanopillar sizes and corresponding phases and transmittances in the circularly polarized detection region, where $\mathbf{T}_{x^\prime}$ and $\mathbf{T}_{y^\prime}$ are transmission coefficient of lights with polarization along $x^\prime$ and $y^\prime$, respectively.}
\label{Tab:pillarsize}
\end{table}

\subsection*{B. Numerical simulation of the designed metasurface}
We simulate the performance of designed metasurface employing finite-difference time-domain (FDTD) method. Limited by computational memory space, we scale down the size of metasurface to $21\mu$m$\times21\mu$m and perform the simulation. The distribution of power intensity on the focal plane with input polarizations of $\ket{H}, \ket{V}, \ket{+}, \ket{-}, \ket{R}$ and $\ket{L}$ are shown in Supplementary Fig.~\ref{Fig:metasimulation2}(a)-(f). It can be seen that the six incident polarizations can be split and focused at the designed positions on the focal plane. As expected, for a specific polarization, the field intensity of its orthogonal polarization on the focal plane is almost zero, while the field intensity of the other two groups of states is half of the total field intensity. Here, the focused field intensity is defined as the total energy within a circle centered at the focal spot, with a radius of 1.5 times the full width at half-maximum (FWHM)~\cite{balli2020hybrid}. Arbitrary polarization state can be represented by the Stokes vector formalized as $\bm{S}=[S_0,S_1,S_2,S_3]^{T}$. The elements $S_i$ are defined by 
\begin{equation}
\begin{split}
&S_0=I_H+I_V=I_++I_-=I_R+I_L,\\
&S_1=I_H-I_V,\\
&S_2=I_+-I_-,\\
&S_3=I_R-I_L, 
\end{split}
\end{equation}
where $I_p$ is the power intensity of polarization component $p\in[H, V, +, -, R, L]$. The results of normalized Stokes parameter $\bm{s}=[s_1, s_2, s_3]$ with
\begin{equation}
\begin{split}
&s_1=\frac{I_H-I_V}{I_H+I_V},\\
&s_2=\frac{I_+-I_-}{I_++I_-},\\
&s_3=\frac{I_R-I_L}{I_R+I_L},  
\end{split}
\end{equation} 
are shown in Supplementary Fig.~\ref{Fig:metasimulation2}(g)-(h).
\begin{figure*}[h!tb]%
\centering
\includegraphics[width=0.9\textwidth]{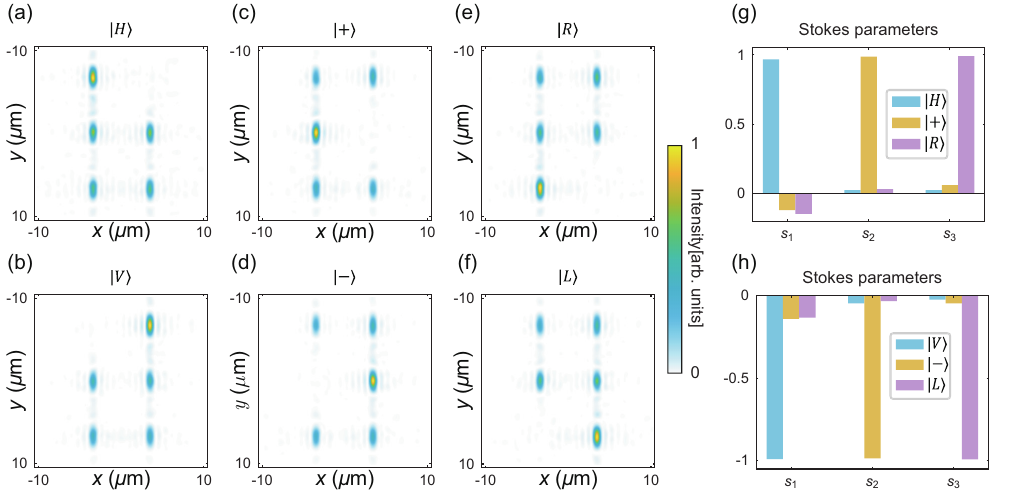}
\caption{\textbf{Numerical simulation of power intensity on the focal plane and reconstructed Stokes parameters.} \textbf{(a)}-\textbf{(f),} Spatial distribution of light intensity at the focal plane with input polarizations of$\ket{H}$,$\ket{V}$, $\ket{+}$, $\ket{-}$, $\ket{R}$ and $\ket{L}$. \textbf{(g)}-\textbf{(h),} The reconstructed Stokes parameters of the input polarizations of $\ket{H}$,$\ket{V}$, $\ket{+}$, $\ket{-}$, $\ket{R}$ and $\ket{L}$.}
\label{Fig:metasimulation2} 
\end{figure*}

Compared to the ideal values, the average error of reconstructed Stokes parameters $s_1$, $s_2$ and $s_3$ are 0.097, 0.027 and 0.029 respectively, where the errors in $\ket{H}/\ket{V}$ basis is larger than that in $\ket{+}/\ket{-}$ and $\ket{R}/\ket{L}$ basis. This is mainly caused by the asymmetric response of region 1 with input polarization of $\ket{H}$ and $\ket{V}$. To verify this, we simulate the optical response of region 1 with input polarization of $\ket{H}$ and $\ket{V}$, respectively. The simulation is performed within range $x\in[-3\mu m, 3\mu m]$ and $y\in[-1\mu m, 1\mu m]$, which is scaling-down of region 1. As shown in Supplementary Fig.~\ref{Fig:metasimulation1}(a), the optical responses with input polarization of $\ket{H}$ and $\ket{V}$ are asymmetric with respect of $y$ axis, leading to different transmit efficiency on focal plane. This is verified by simulation of distribution of power intensity on focal plane with input polarization of $\ket{+}=\frac{1}{\sqrt{2}}(\ket{H}+\ket{V})$ as shown in Supplementary Fig.~\ref{Fig:metasimulation1}(d), where the output power intensities are unbalanced and introduces more errors in reconstruction of $s_1=\frac{I_H-I_V}{I_H+I_V}$. 

In contrast to region 1, region 2~($\ket{+}$/$\ket{-}$ section) and region 3~($\ket{R}$/$\ket{L}$ section) response their input polarization in symmetric manner. As shown in Supplementary Fig.~\ref{Fig:metasimulation1}(b) and (c), the distributions of response intensity with input polarization of $\ket{+}$~($\ket{R}$) and $\ket{-}$~($\ket{L}$) are symmetric with respect to $x=0$, which consequently leads the balanced splitting of power intensity of input polarization $\ket{H}$ as shown in Supplementary Fig.~\ref{Fig:metasimulation1}(e) and (f). Therefore, the errors in reconstruction of $s_2=\frac{I_+-I_-}{I_++I_-}$ and $s_3=\frac{I_R-I_L}{I_R+I_L}$ are smaller than that of $s_1$.

\begin{figure}[h!tb]%
\centering
\includegraphics[width=0.9\textwidth]{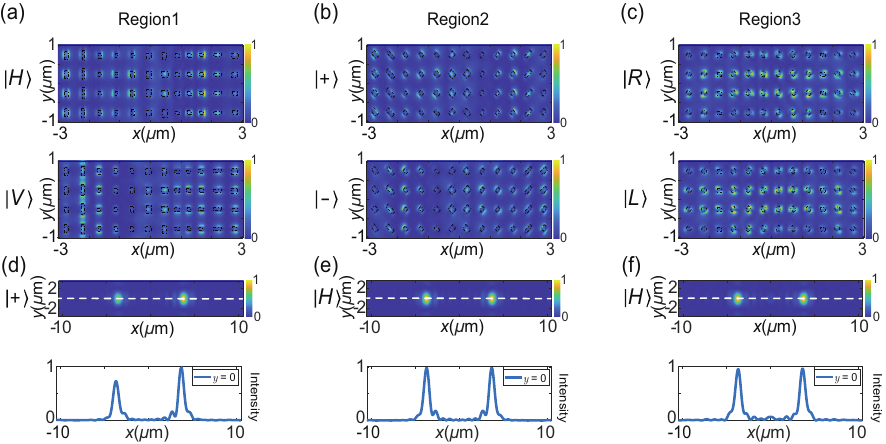}
\caption{\textbf{Simulation of optical response of metasurface with different input polarizations.} Optical response with input polarization of \textbf{(a),} $\ket{H}$ and $\ket{V}$, \textbf{(b),} $\ket{+}$ and $\ket{-}$, \textbf{(c),} $\ket{R}$ and $\ket{L}$. The distribution of power intensity on focal plane with input polarization of \textbf{(d),} $\ket{+}$, \textbf{(e),} $\ket{H}$ and \textbf{(f),} $\ket{H}$. The power intensity are represented in arbitray unit.}
\label{Fig:metasimulation1}
\end{figure}

Discretization would also introduce optical loss. For example, different arrangements of nanopillars in three regions would introduce a ``cut-off" in phase configuration on metasurface. We simulate the phase configuration on metasurface with input polarization of $\ket{V}$. As shown in Supplementary Fig.~\ref{Fig:metasimulation}(a), there are two cut-off lines between three regions, which introduces undesired scattering and consequently increases the optical loss. However, this optical loss is small as most of the incident light is focused around the desired spot as shown in Supplementary Fig.~\ref{Fig:metasimulation}(b) and (c), which is coupled into optical fibers in our experiment.
\begin{figure}[h!tb]%
\centering
\includegraphics[width=0.6\textwidth]{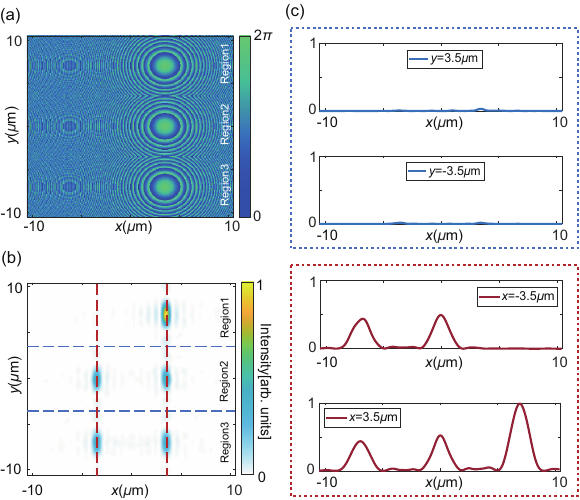}
\caption{\textbf{Simulation of phase configuration and distribution of intensity on focal plane with input polarization of $\ket{V}$. (a),} The phase distribution on metasurface. \textbf{(b),} The distribution of intensity on the focal plane. \textbf{(c),} The distribution of intensity along the dashed lines in \textbf{(b)}.}
\label{Fig:metasimulation}
\end{figure}

\subsection*{C. Experimental reconstruction of Stokes parameters of polarization with metasurface}
The setup to reconstruct the Stokes parameters is shown in Supplementary Fig.~\ref{Fig:metacharact1}. The wavelength of the incident light is 810nm followed by combination of a linear polarizer and a quarter-waveplate~(QWP), which manipulates the polarization of incident light. Then, the transmitted light focused on the focal plane is captured by a $20 \times$ objective lens~(OL) and recorded on a CMOS image sensor. We test our metasurface with six input polarizations $p\in[H, V, +, -, R, L]$, and record the distribution of power intensity on the focal plane for each input. According to the power intensity, we calculate the normalized Stokes parameter $\bm{s}=[s_1, s_2, s_3]$, and the results are shown in Figure~1 in main text.

\begin{figure}[h!tb]%
\centering
\includegraphics[width=0.6\textwidth]{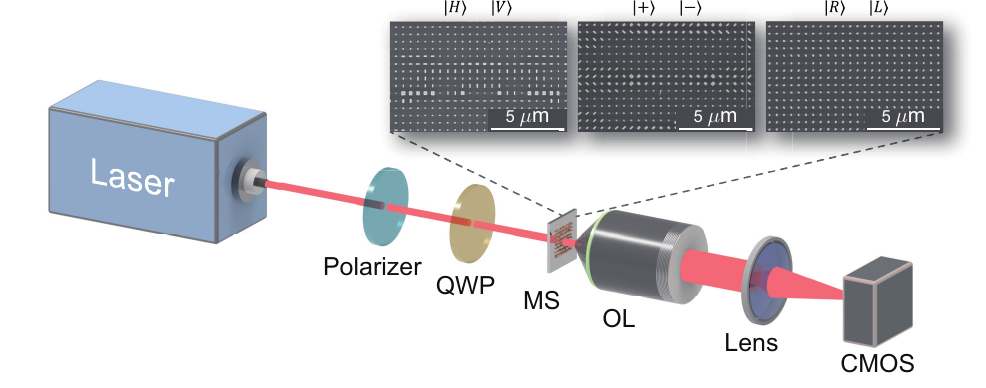}
\caption{Experimental setup to reconstruction of Stokes parameters of polarization with metasurface. HWP: half-wave plate. QWP: quarter-wave plate. MS: metasurface. OL: objective lens. CMOS: complementary metal-oxide-semiconductor.}
\label{Fig:metacharact1}
\end{figure}

\subsection*{D. Imperfections in metasurface}
As shown in main text, the experimental results of reconstruction of Stokes parameters, shadow tomography estimation and MLE reconstruction conclude that the metasurface introduces measurement errors from 0.07 to 0.1. The errors are mainly attributed to the limitations in design of metasurface and the imperfections in fabrication of metasurface. 

The key ingredient in the design of metasurface is to discretize the phase front. According to requirement of deflecting individual polarization to desired direction, we calculate a discrete phase accumulation of metasurface. Then, a periodic array of nanopillars is designed to realize such a discrete phase accumulation. However, the main limitation in such design is discretization itself, which inevitably introduces polarization measurement errors in our case. The discretization of phase front is an approximation of its continuous counterpart~(bulk optics), which limits the accuracy of phase modulation. Consequently, the metasurface to deflect input polarization (for example $\ket{H}$) to desired direction cannot completely block its orthogonal polarization~($\ket{V}$) transmitting along the same direction, which introduces polarization measurement error. Particularly, to separate circular polarizations, i.e., $\ket{R}$ and $\ket{L}$, the cross-polarization effect has been employed in the design of metasurface, in which the conversion between two orthogonal circular polarizations~($\ket{R}\to\ket{L}, \ket{L}\to\ket{R}$) is firstly taken place on the metasurface. However, it is still challenging to realize complete conversion between $\ket{R}$ and $\ket{L}$ on metasurface. Specifically, the aspect ratio constraints of nanopillars, adjacent coupling between nanopillars as well as material absorption would reduce the efficiency of conversion, which consequently increases the errors in polarization measurement. 

On the other hand, the metasurface to realize discrete phase modulation can be considered as a grating. For large bending angles~($43^\circ$ in our design), it inevitably deflects the incident polarization into other grating orders~(undesired directions)~\cite{Sell2017NL}.

Indeed, there are several schemes and techniques can improve the performance of metasurface.

\noindent \textbf{High-order diffraction suppression.} The high-order diffraction caused by the beam deflection can be suppressed by design of metasurface. For example, asymmetric grating profile~\cite{Sell2017NL} and nonperiodic metagrating designs~\cite{Born2023NC} have been proposed to suppress high-order diffraction and deflect the light into a single desired order.

\noindent \textbf{Multi-layer metasurface.} Stacking multiple layers of metasurface with varied polarization filtering functionalities is able to enhance the accuracy of polarization control. For example, utilizing the double-layer chiral metasurface~\cite{Basiri2019Light}, the average measurement errors of Stokes parameters $s_1$, $s_2$, and $s_3$ were achieved at near-infrared wavelengths of 1.9\%, 2.7\% and 7.2\%, respectively. The corresponding results in our work are 10.1\%$\pm$0.5\%, 8.6\%$\pm$0.5\% and 7.3\%$\pm$0.5\%.
   
\noindent \textbf{Calibration.} For general applications, the calibration process could eliminate errors introduced by inhomogeneity of the incident light on metasurface, including power intensity and incident angle~\cite{Tang2018}. Particularly, it has been proved that after calibration, the performance of metasurface-enabled polarimeter is comparable to that of bulk optics~\cite{Li2023}.

\section*{Supplementary Note 5: Realizing projection on $\ket{\psi_l}$ with equal probability}
It is impossible to equally split input light into three regions due to the mode mismatch between incident light~(Gaussian beam) and metasurface~(square). Instead, we post select the photons passing through each region with equal probability. First, the beam waist of the input light is carefully adjusted to be $w_0=\sqrt{2}\times 210~\mu$m with lens, which enables the maximal overlap between beam waist and metasurface~(210~$\mu$m square). Then, we carefully locate the metasurface at the center of beam waist, which enables the equal probability of a single photon passing through region 1~($\ket{H}/\ket{V}$ basis) and region 3~($\ket{R}/\ket{L}$ basis), i.e., the count rates of collected photons passing through these two regions are the same. However, the single photon passes through region 2~($\ket{+}/\ket{-}$ basis) with higher probability due to the nature of Gaussian distribution. In our experiment, we randomly discard the collected photons passing through region 2 to make the count rate equal to that of region 1 and region 3. Such experimental setting enables the equal probability of single photon passing through three regions.

\end{document}